%% file: main.tex
\documentclass[sigconf, nonacm]{acmart}

\newcommand\vldbavailabilityurl{https://github.com/purduedb/iPDB}
\newcommand\vldbpagestyle{plain} 

\usepackage{numprint} 
\usepackage{array}
\usepackage{listings}
\usepackage{xspace}
\usepackage[table, dvipsnames]{xcolor}
\usepackage{cleveref}
\usepackage{multirow}
\usepackage{subcaption}
\usepackage{makecell}
\usepackage{fancyhdr}
\usepackage{booktabs}
\usepackage{array}
\usepackage{tikz}
\usepackage{threeparttable}
\usepackage{amsmath}
\usepackage[usestackEOL]{stackengine}
\usepackage{pgfplots, pgfplotstable}

\usepackage{amssymb}
\usepackage{pifont}
\usepackage{enumitem}
\usepackage{rotating} 
\usepackage{graphicx} 
\usepackage[most]{tcolorbox}
\usepackage{xcolor}

\begin{document}

\newcommand{\dbname}[0]{iPDB}
\newcommand{\improvement}[0]{2.5x mean speedup, with speedups of up to 30x}
\title{\dbname{} - Optimizing Semantic SQL Queries}

\author{Udesh Kumarasinghe}
\email{ukumaras@purdue.edu}
\affiliation{%
  \institution{Purdue University}
  \streetaddress{ 305 N University St.}
  \city{West Lafayette}
  \state{IN}
  \country{USA}
  \postcode{47907}
}
\author{Tyler Liu}
\email{liu3450@purdue.edu}
\affiliation{%
  \institution{Purdue University}
  \streetaddress{ 305 N University St.}
  \city{West Lafayette}
  \state{IN}
  \country{USA}
  \postcode{47907}
}

\author{Ahmed R. Mahmood}
\email{amahmoo@google.com}
\affiliation{%
  \institution{Google Inc.}
  \streetaddress{}
  \city{{Mountain View}}
  \state{CA}
  \country{USA}
}

\author{Chunwei Liu}
\email{chunwei@purdue.edu}
\affiliation{%
  \institution{Purdue University}
  \streetaddress{ 305 N University St.}
  \city{West Lafayette}
  \state{IN}
  \country{USA}
  \postcode{47907}
}

\author{Walid G. Aref}
\email{aref@purdue.edu}
\affiliation{%
  \institution{Purdue University}
  \streetaddress{ 305 N University St.}
  \city{West Lafayette}
  \state{IN}
  \country{USA}
  \postcode{47907}
}
\renewcommand{\shortauthors}{Kumarasinghe et al.}

\begin{abstract}
\input{0_abstract}
\end{abstract}

\input{setting}

\maketitle

\pagestyle{\vldbpagestyle}
\begingroup\small\noindent\raggedright\textbf{Keywords:}\\
Semantic Operators, Query Optimization, LLM in SQL, DB4AI
\endgroup
\ifdefempty{\vldbavailabilityurl}{}{
\vspace{.3cm}\\
\begingroup\small\noindent\raggedright\textbf{Code Availability:}\\
The source code have been made available at \url{\vldbavailabilityurl}.
\endgroup
}

\section{Introduction}
\label{sec:introduction}
\input{1_introduction}

\section{Semantic SQL Queries using ML and LLMs}
\label{sec:syntax_definition}
\input{3_syntax_definition}

\section{System Overview and Design}
\label{sec:overview}
\input{4_overview}

\section{Implementation of the {\tt PREDICT} Operator}
\label{sec:phy_operator}
\input{5_phy_operator}

\section{Optimizing Semantic SQL Queries}
\label{sec:optim}
\input{6_optim}

\section{Performance Evaluation}
\label{sec:evaluation}
\input{8_evaluation}

\section{Related Work}
\label{sec:related_work}
\input{9_related_work}

\section{Conclusion}
\label{sec:conclusion}
\input{10_conclusion}

\bibliographystyle{ACM-Reference-Format}
\bibliography{reference}

\end{document}

%% file: 0_abstract.tex
The Structured Query Language (SQL) has remained the standard query language for databases. SQL is highly optimized for processing structured data laid out in relations. Meanwhile, in the present application development landscape, it is highly desirable to utilize the power of learned models to perform complex tasks. Large language models (LLMs) have been shown to understand and extract information from unstructured textual data. However, SQL as a query language and accompanying relational database systems are either incompatible or inefficient for workloads that require leveraging learned models. This results in complex engineering and multiple data migration operations that move data between the data sources and the model inference platform. This paper presents \dbname{}, a relational data system that supports in-database machine learning (ML) and large language model (LLM) inferencing using extended SQL syntax. In \dbname{}, LLMs and ML calls can function as semantic projects, as predicates to perform semantic selects and semantic joins, or for semantic grouping in group-by clauses. \dbname{} has a new relational predict operator along with semantic query optimizations that enable users to write and efficiently execute semantic SQL queries, outperforming other state-of-the-art systems {by \improvement{}}. 

%% file: setting.tex
\definecolor{codegreen}{rgb}{0.228,0.695,0.195}
\definecolor{codepink}{rgb}{0.7539,0.1992,0.4531}
\definecolor{codeblue}{rgb}{0.1914,0.457,0.707}
\definecolor{codegray}{rgb}{0.59,0.59,0.59}
\definecolor{codeyellow}{rgb}{0.9254, 0.5529, 0.366666667}
\definecolor{backcolour}{rgb}{1,1,1}

\pgfplotsset{compat=1.18}

\newcommand{\EvalHeader}{
  \toprule
    & \multirow{2}{*}{\textbf{Model}} & \multicolumn{3}{c|}{\textbf{(B1) LOTUS}} & \multicolumn{3}{c|}{\textbf{(B2) Flock}}
    & \multicolumn{3}{c|}{\textbf{(B3) BigQuery}} & \multicolumn{3}{c}{\textbf{(B4) \dbname{}}} \\
    \cmidrule{3-5} \cmidrule(lr){6-8} \cmidrule(lr){9-11} \cmidrule(lr){12-14}
    & & \textbf{Calls|Cost} & \textbf{Quality} & \textbf{Latency}
    & \textbf{Calls|Cost} & \textbf{Quality} & \textbf{Latency}
    & \textbf{Calls|Cost} & \textbf{Quality} & \textbf{Latency}
    & \textbf{Calls|Cost} & \textbf{Quality} & \textbf{Latency} \\
  \midrule
}

\newcommand{\ScenarioRow}[1]{%
  \multicolumn{14}{c}{\textbf{#1}}\\
}

\newcommand{\ContinuedNote}{%
  \multicolumn{13}{r}{\small\itshape Table~\ref{tab:experimental_results_all} continued from previous page.}\\[-0.3ex]
}

\newcommand{\cmark}{\ding{51}}%
\newcommand{\xmark}{\ding{53}}%
\newcommand{\clockmark}{\ding{67}}%

\def\arraystretchresulttable{0.8}


\newcommand*{\eg}{e.g.,\@\xspace}
\newcommand*{\ie}{i.e.,\@\xspace}
\newcommand*{\cf}{cf.\@\xspace}
\newcommand*{\etal}{~et~al.\@\xspace}
\newcommand*{\dash}{\textemdash\@\xspace}
\newcommand*{\etc}{~etc.\@\xspace}

\newcommand*{\s}{\,s\@\xspace}
\newcommand*{\ms}{\,ms\@\xspace}
\newcommand*{\GHz}{\,GHz\@\xspace}
\newcommand*{\B}{\,B\@\xspace}
\newcommand*{\KB}{\,KB\@\xspace}
\newcommand*{\MB}{\,MB\@\xspace}
\newcommand*{\GB}{\,GB\@\xspace}
\newcommand*{\TB}{\,TB\@\xspace}
\newcommand*{\percent}{\,\%\@\xspace}
\newcommand*{\M}{\,M\@\xspace}

\clubpenalty = 10000
\widowpenalty = 10000

\newcommand\paragraphnospace[1]{\noindent{\bfseries#1\space}}
\newcolumntype{C}[1]{>{\centering\arraybackslash}p{#1}}

\newcommand{\watermarked}[2]{
    \begin{tikzpicture}
        \node[opacity=1.0] at (0,0) {#1};
        \node[rotate=45, color=red, font=\Huge, opacity=0.4] at (0,0) {#2};
    \end{tikzpicture}
}
\newcommand{\draftwatermarked}[1]{\watermarked{#1}{DRAFT}}

\definecolor{dkgreen}{rgb}{0,0.6,0}
\definecolor{gray}{rgb}{0.5,0.5,0.5}
\definecolor{mauve}{rgb}{0.58,0,0.82}

\newcommand*\circled[1]{\tikz[baseline=(char.base)]{
            \node[shape=circle,draw,fill=white,inner sep=2pt] (char) {#1};}}

\lstdefinelanguage{SQL}{
    morekeywords={CREATE, MODEL, ON, PROMPT, PATH, API, SELECT, FROM, WHERE, HAVING, ORDER, GROUP, BY, DESC, ASC, AS, LLM, AGG, PROMPT, OPTIONS, INNER, NATURAL, TABULAR, TABLE, FEATURES, OUTPUT, JOIN, AND, OR},
    morekeywords={INTEGER, VARCHAR, BOOLEAN},
    sensitive=true, 
    morecomment=[l]{--}, 
    morecomment=[s]{/*}{*/}, 
    moredelim=[s][\color{codepink}]{\{\{}{\}\}}, 
    morestring=**[d]{'},
    morestring=[d]{\\'},
    morestring=**[d]{"}
} %

\lstdefinestyle{SQLStyle}{
    aboveskip=3mm,
    belowskip=3mm,
    backgroundcolor=\color{backcolour},   
    keywordstyle=\color{codegreen}\bfseries,
    numberstyle=\tiny\color{codegray},
    stringstyle=\color{codeblue},
    basicstyle=\ttfamily\footnotesize,
    breakatwhitespace=false,         
    breaklines=true,                 
    captionpos=b,                    
    keepspaces=true,                 
    numbers=left,                    
    numbersep=5pt,                  
    showspaces=false,                
    showstringspaces=false,
    showtabs=false,                  
    tabsize=2
}

\lstdefinestyle{SQLStyleTiny}{
    backgroundcolor=\color{backcolour},   
    keywordstyle=\color{codegreen}\bfseries,
    numberstyle=\tiny\color{codegray},
    commentstyle=\color{codeyellow}\bfseries,
    stringstyle=\color{codeblue},
    basicstyle=\ttfamily\scriptsize,
    breakatwhitespace=false,         
    breaklines=true,                 
    captionpos=b,                    
    keepspaces=true,                 
    numbers=none,             
    comment=[l]{--},
    showspaces=false,                
    showstringspaces=false,
    showtabs=false,                  
    tabsize=2
}

\lstset{style=SQLStyle}

\usetikzlibrary{patterns, 
    fit,
    patterns.meta,
    backgrounds,
    shapes.geometric, 
    arrows.meta, 
    positioning,
    calc
    }

\tikzset{
    fancycellbase/.style={
            rounded corners=2pt,
            inner sep=2pt,
            text height=1.3ex,
            text depth=.0ex,
            text centered,
            text=black
    }}
    
\tikzset{
    fancycellhatchbase/.style={
            opacity=0.5,
            rounded corners=2pt,
            fit=(X),
            inner sep=-0.1pt,
    }}

\tikzset{
    fancycellhatch1/.style={
            fancycellhatchbase,
            opacity=0.7,
            pattern={
                Lines[angle=-45,
                distance={3pt}, 
                line width=1pt]},
            pattern color = white,
    }
}

\tikzset{
    fancycellhatch2/.style={
            fancycellhatchbase,
            opacity=0.7,
            pattern={
                Lines[angle=90,
                distance={3pt}, 
                line width=1pt]},
            pattern color = white,
    }
}

\tikzset{
    fancycellhatch3/.style={
            fancycellhatchbase,
            opacity=0.7,
            pattern={
                Dots[angle=90,
                distance={3pt}, 
                radius=0.8pt]},
            pattern color = white,
    }
}

\tikzset{
    fancycellhatch4/.style={
            fancycellhatchbase,
            opacity=0.7,
            pattern={
                Hatch[angle=45,
                distance={6pt}, 
                line width=0.7pt]},
            pattern color = white,
    }
}

\newcommand{\fancycellC}[5]{%
  \begingroup
    \definecolor{__tmpfill}{HTML}{#2}%
    \begin{tikzpicture}[baseline=(X.base)]
        \node[fancycellbase,
            minimum width=#1,
        ] (X) {#3};
        \begin{scope}[on background layer]
            \node[fancycellhatchbase, fill=__tmpfill!80]{};
            \node[#4] {}; 
        \end{scope}
    \end{tikzpicture}%
  \endgroup
}

\newcommand{\fancycellQ}[3]{%
  \begingroup
    \definecolor{__tmpfill}{HTML}{#2}%
    \begin{tikzpicture}[baseline=(X.base)]
        \node[fancycellbase,
            minimum width=#1,
        ] (X) {#3};
        \begin{scope}[on background layer]
            \node[fancycellhatchbase, fill=__tmpfill]{};
        \end{scope}
    \end{tikzpicture}%
  \endgroup
}

\newcommand{\fancycell}[3][3.6em]{\fancycellC{#1}{#2}{#3}{fancycellhatch3}{ }}

\newcommand{\fancycellGreen}[1]{\fancycellC{3.8em}{7ede94}{\textbf{#1}}{fancycellhatch1}{ $\checkmark$ }}
\newcommand{\fancycellYellow}[1]{\fancycellC{3.8em}{f8c64b}{#1}{fancycellhatch2}{ $\blacktriangle$ }}
\newcommand{\fancycellGray}[1]{\fancycellC{13em}{9ca5ad}{#1}{fancycellhatch3}{ $\bullet$ }}
\newcommand{\fancyCellRed}[1]{\fancycellC{0em}{f84b7a}{ \xmark }{fancycellhatchbase}{}}
\newcommand{\fancycellYellowX}[1]{\fancycellC{0em}{f8c64b}{\xmark}{fancycellhatch2}{ $\blacktriangle$ }}

\newcommand{\fancycellSelect}[1]{\fancycellQ{0.1em}{2563eb}{#1}}
\newcommand{\fancycellProject}[1]{\fancycellQ{0.1em}{facc15}{#1}}
\newcommand{\fancycellJoin}[1]{\fancycellQ{0.1em}{f43f1a}{#1}}
\newcommand{\fancycellRelation}[1]{\fancycellQ{0.1em}{2ecc71}{#1}}
\newcommand{\fancycellAgg}[1]{\fancycellQ{0.1em}{9ed3dc}{#1}}
\newcommand{\fancycellEmbed}[1]{\fancycellQ{0.1em}{fcb7c7}{#1}}

\newcommand{\fcmark}[0]{\fancycellQ{0.1em}{7ede94}{\cmark}}
\newcommand{\fxmark}[0]{\fancycellQ{0.1em}{f84b7a}{\xmark}}
\newcommand{\fwmark}[0]{\fancycellQ{0.1em}{f8c64b}{\cmark}}

\newcommand{\fancycellNormal}[1]{\fancycellC{3.8em}{dddddd}{#1}{fancycellhatchbase}{  }}

\newcommand{\fancycellnormal}[3][3.6em]{%
  \begingroup
    \begin{tikzpicture}[baseline=(X.base)]
        \node[fancycellbase,
            minimum width=#1,
        ] (X) {#3};
        \node[] (Y) at (X.east){};
        \begin{scope}[on background layer]
            \node[fancycellhatchbase, fill=#2]{};
        \end{scope}
    \end{tikzpicture}%
  \endgroup
}

%% file: 1_introduction.tex
Recent advances in AI have opened the way for breakthroughs in applications, e.g., question answering~\cite{qa_survey_2025}, information retrieval~\cite{info_ret_2024}, and language understanding~\cite{llm_survey_2025}. However, Structured Query Language (SQL) has remained the standard and commonly used query language for databases. Although SQL is optimized for querying tables with structured attributes, e.g., numeric values and  strings, it is inefficient on workloads that require semantic reasoning and extracting semantic information from unstructured data, which is highly relevant in today's application development landscape.

\input{tables/features.tex}

Having learned model inference from within SQL to run on the data  in the database enables a wide array of tasks, e.g., classification, sentiment analysis, summarization, and text generation. Furthermore, model inference capabilities within SQL enables users to leverage traditional SQL functionality and expressiveness to create more complex semantic queries that cannot be accomplished by simply feeding rows of unprocessed data into models. Additionally, by identifying patterns in semantic queries and performing logical and physical optimizations, it is possible to efficiently execute large-scale inference tasks. These optimizations are vital as LLM workloads are expensive in both runtime and monetary terms~\cite{palimpzestCIDR}. 
 
Several systems have been proposed that extract and filter information from  documents~\cite{lotus_2024,palimpzestCIDR,docetl_2025}. Although these systems demonstrate the potential and applications of semantic query processing, they either introduce an imperative query language,  only support unstructured data, or lack optimizations for relational data. Table~\ref{tbl:feat_comp} gives a comparison of these systems. To address this gap, we extend standard SQL queries by introducing first-class prediction functions. Specifically, we augment SQL with ML model prediction and LLM inference. These added functions enable context-aware predictions and semantic operations, e.g., semantic selects, projects, joins, aggregates, and relations, surpassing the capabilities of traditional relational operations and predicates. For example, given a database table of products where each tuple is a PC component, we can find the list of compatible motherboards and CPUs, by self joining with an natural language condition. The following Listing~\ref{lst:intro_example} shows the extended SQL syntax for this query.

\begin{lstlisting}[caption={This semantic SQL query performs LLM-based semantic join on column \emph{name} of motherboards and CPUs},label={lst:intro_example},language = SQL]
SELECT c.name, m.name
FROM Product AS m JOIN Product AS c 
ON LLM o4mini (PROMPT 'is CPU  {{c.name}} {compatible BOOLEAN} with motherboard {{m.name}}')
WHERE m.category = 'Motherboard' AND c.category = 'CPU';
\end{lstlisting}


Although supporting semantic queries in a relational database system is important, realizing an efficient system is non-trivial. We observe the following challenges in implementing this system.

\begin{itemize}
    \item Inference in learned models is expensive and has a higher latency than traditional database operations. Thus, inferencing will be a bottleneck in semantic queries, necessitating prioritizing it when optimizing semantic queries.
    \item {A wide range of learned models and execution environments exist. Users are required to setup and configure each model imperatively without a declarative inference syntax.}
    \item For queries that use LLMs, schema alignment must be ensured to fit text-native LLMs into schema-driven relational databases (i.e., typed columns and constraints). The inference operator must ensure that LLM output can be parsed into consistent column types (e.g., VARCHAR, INT, DOUBLE).
    \item Learned models may produce missing values, values with type mismatches, or hallucinate outputs. The system must minimize the possibility of errors and handle adversities. 
    \item Inferencing is prone to runtime errors as a result of external factors e.g., memory overflow, GPU latency, and network errors. The system must handle these errors gracefully.
\end{itemize}


To overcome these challenges, we realize \dbname{}, a data system with extended SQL and execution engine that support semantic queries. \dbname{} treats learned models as first-class citizens, where users can define, manage, and infer within the clauses of SQL, i.e., \verb|SELECT|, \verb|FROM|, \verb|WHERE|, \verb|GROUP BY|, and \verb|ORDER BY|. We utilize DuckDB~\cite{duckdb_2019}, a columnar-based relational engine, to reuse its parser, relational query primitives, storage, and base query engine. To process \dbname{'s} semantic queries, we extend DuckDB's parser, planner, optimizer, and execution engine. The parser is extended with model manipulation language and predict/LLM clauses to support semantic queries. Model management capabilities are realized by introducing a new system catalog to store and retrieve model metadata. The inference of learned models is achieved through a new physical {\em predict} operator that manages input tuple processing and model probing, and extracts output attributes for schema compliance. New optimizations are introduced within the predict operator and the plan optimizer to minimize the latency of model inference.


The following are the key contributions of the \dbname{} system:

\begin{itemize}
\item {Extended SQL Syntax for Model Inference: New {\tt PREDICT} and an {\tt LLM} SQL clauses are introduced to enable semantic projections, semantic table generation, semantic predicates in the where clause, semantic grouping, and in order-by clauses. These operators integrate model inference into Relational Algebra to take tables as input, perform model probing, and produce schema-compliant output tables.}
\item {New optimizations for ML model and LLM inference, e.g., input deduplication, multi-row marshaling, semantic select pulling, semantic join vs. traditional join, semantic join decomposition and cost-based semantic operator ordering.}
\item A unified system that extends all system layers to integrate model inference and metadata management, enabling the execution of relational and semantic query processing within a single declarative SQL framework.
\item {Experimental study shows that \dbname{} with the optimizations achieves \improvement{} over state-of-the-art systems by reducing the number of tokens and model calls while preserving answer quality.}
\end{itemize}

The rest of the paper proceeds as follows. 
Section~\ref{sec:syntax_definition} presents \dbname{'s} extended SQL for semantic queries. Section~\ref{sec:overview} overviews \dbname{} and its major components. Section~\ref{sec:phy_operator} presents the design and realization of the \verb|PREDICT| operator. Sections~\ref{sec:optim} and~\ref{sec:evaluation} present \dbname{'s} optimizations and performance results, respectively. Section~\ref{sec:related_work} discusses related work, and Section~\ref{sec:conclusion} concludes the paper.

%% file: tables/features.tex
\begin{table*}[h]
\centering
\footnotesize
\begin{tabular}{p{2.2cm}|p{2cm}|p{2cm}|p{2.2cm}|p{2.3cm}|p{3.5cm}}
\toprule
\textbf{Feature} & \textbf{LOTUS} & \textbf{Flock} & \textbf{BigQuery} & \textbf{pz} & \textbf{The Proposed \dbname{} System} \\
\midrule
Declarative Syntax & \fxmark & \fxmark & \fcmark & \fxmark & \fcmark \\
\midrule
Structured Outputs & \fwmark~ Untyped & \fcmark~ JSON Schema & \fcmark~ In Prompt Schema & \fcmark ~Dictionary & \fcmark~ In Prompt Schema \\
\cmidrule{2-6}
Multi Modal & \fcmark ~Text, Image & \fxmark ~Text & \fcmark ~Text, Image & \fcmark ~Text, Image, Audio & \fcmark Text, ~Image \\
\midrule
Supported Operators & \fancycellProject{$\pi^{s}$} \fancycellSelect{$\sigma^{s}$} \fancycellJoin{$\bowtie^s$}  \fancycellAgg{$\gamma^s$} & \fancycellProject{$\pi^{s}$} \fancycellSelect{$\sigma^{s}$} \fancycellAgg{$\gamma^s$} & \fancycellProject{$\pi^{s}$} \fancycellSelect{$\sigma^{s}$} \fancycellJoin{$\bowtie^s$} \fancycellAgg{$\gamma^s$} & \fancycellProject{$\pi^{s}$} \fancycellSelect{$\sigma^{s}$} \fancycellJoin{$\bowtie^s$} \fancycellAgg{$\gamma^s$} & \fancycellProject{$\pi^{s}$} \fancycellSelect{$\sigma^{s}$} \fancycellJoin{$\bowtie^s$} \fancycellAgg{$\gamma^s$} \fancycellRelation{$\rho^s$} \\
\cmidrule{2-6}
Table functions & \fcmark & \fxmark & \fcmark & \fcmark & \fcmark \\
\cmidrule{2-6} 
ML support & \fxmark & \fxmark & \fcmark & \fxmark & \fcmark \\
\cmidrule{2-6} 
Supported LLMs & \fcmark API & \fcmark~ API, ollama & \fwmark ~Gemini Only & \fcmark ~API & \fcmark~ API, Local \\
\cmidrule{2-6} 
In DB inference & \fxmark & \fxmark & \fxmark & \fxmark & \fcmark \\
\cmidrule{2-6}
Exception handling & \fxmark ~On LLM error & \fxmark ~On LLM error & \fcmark & \fcmark & \fcmark \\
\midrule
Logical\newline Optimizations & \fxmark & \fxmark & \fcmark & \fcmark~ Logical rewrites & \fcmark~ Semantic Select Pulling, Semantic Select vs Join Ordering, Cost-based Ordering, Semantic Join Decomposition \\
\cmidrule{2-6}
Physical\newline Optimizations & \fxmark & \fxmark & \fcmark~\verb|LIMIT| early stop & \fcmark ~Model selection & \fcmark~ \verb|LIMIT| early stop \\
\cmidrule{2-6}
Intra Operator\newline Optimizations & \fcmark~ Parallelization & \fcmark~ Static Row Marshaling & \fcmark & \fcmark ~Static Row Marshaling, Parallelization & \fcmark~ Deduplication, Dynamic Row Marshaling, Model Aware Input Ordering, Parallelization, Null Handling, Semantic Operator Merging \\
\bottomrule
\end{tabular}
\caption{Comparison of existing LLM-enabled data systems (LOTUS, Flock, BigQuery, and pz) with the proposed \dbname{} across supported features, multi-modal capabilities, LLM compatibility, and optimization techniques. 
Semantic Operators: \protect\fancycellProject{$\pi^{s}$- Semantic Project}, \protect\fancycellSelect{$\sigma^{s}$ - Semantic Select}, \protect\fancycellJoin{$\bowtie^s$ - Semantic Join}, \protect\fancycellAgg{$\gamma^s$ - Semantic Aggregation}, and \protect\fancycellRelation{$\rho^s$ - Semantic Relation}}\label{tbl:feat_comp}.
\end{table*}

%% file: 3_syntax_definition.tex
In this section, we present \dbname{'s} extended SQL for semantic queries. We define a model upload statement, and a generalized {\tt PREDICT} and {\tt LLM} calling expression for inference.

\subsection{Model Upload}

The model-upload command uploads model metadata to \dbname{} that verifies the model and stores the metadata in a model catalog table. The current LLM landscape has multiple open-source and proprietary models. Models may have multiple versions of different sizes with trade-offs in quality, latency, and cost. \dbname{} supports multiple models. It introduces following command for remote model upload:

\begin{lstlisting}[caption={Model Upload Command},label={lst:model_upload_remote},language = SQL]
CREATE LLM MODEL o4mini 
PATH 'o4-mini'
ON PROMPT API 'https://api.openai.com/v1/';
\end{lstlisting}

Argument \texttt{MODEL}  is the model's name (e.g., \texttt{o4mini}) to be used in inference queries,  \texttt{PATH}  is the model identifier (e.g., \verb|'o4-mini'|) in the API, and \texttt{API} is the API's URL (e.g., \verb|'https://api.openai.com'|).

Users can run local LLM models instead of calling an LLM vendor through their API. \dbname{} is compatible with local LLM models that use a built-in executer to load and run them. Local model is uploaded by providing the LLM's local path in the \verb|PATH| parameter and dropping the  \verb|API| parameter. In both examples, the model is defined as \texttt{ON PROMPT}. This informs \dbname{} that the user will provide the LLM's prompt at query time. A prompt clause in the query allows structured placeholders to define input and output column names, and their corresponding types. Allowing prompt and, consequently, input/output columns during query times enables  users to use the same model for different tasks by changing the prompts.

\dbname{} is compatible with pre-trained deep learning and ML models trained for specific tasks. These models accept input features and predict a corresponding set of output features. \dbname{} uses the same \verb|CREATE MODEL| command to specify the model type. Unlike \verb|LLM| models, ML models need the user to specify input and output columns at model upload time, offloading the imperative model call at query time. Users bind the model to a table or specify the model's input columns from different tables.

\color{black}
\subsection{Model Inference}

With the models uploaded to \dbname{'s} model catalog, users can utilize the models in \verb|LLM| or \verb|PREDICT| clauses in the \verb|SELECT| statement. The following is an example of a semantic SQL query:

\begin{lstlisting}[caption={Model Inference Query},label={lst:scalar_inference},language = SQL]
SELECT name, quantity
FROM Product
WHERE LLM o4mini (PROMPT 'get the {vendor VARCHAR} from product {{name}}') = 'Intel';
\end{lstlisting}
\noindent
where the \verb|LLM| clause specifies the \verb|o4mini| model with the prompt given as an argument. In  \verb|PROMPT|, the input column is defined using \verb|{{...}}| while output columns are defined with \verb|{...}| placeholders but with an additional output type (i.e., \verb|{state VARCHAR}|).

\dbname{} supports three execution modes depending on PREDICT's inputs and outputs: (1)~\emph{Table inference}: allows defining multiple output columns to extract multiple attributes from a set of input columns. It produces a predicted table with one or more predicted columns along with the columns of the input relation. In this case, an \verb|LLM| clause will appear in the \verb|FROM| clause. Predicted columns can be used in consecutive SQL or semantic operators, e.g., project or semantic selects. (2)~\emph{Table Generation}: The input relation is omitted from the \verb|LLM| clause for the LLM to generate a new relation. The LLM call behaves as a scan, where the LLM's data values populate a virtual relation. Table generation enables \dbname{} to extract the LLM's learned information, allowing users to declaratively extract structured data from the LLM as a data source. (3)~\emph{Scalar Inference}: The user can include only a single output column in the prompt, resulting in a behavior comparable to a scalar function. With scalar inference, the \verb|LLM| clause may appear in  \verb|SELECT|, \verb|WHERE|, \verb|ORDER BY|, or \verb|GROUP BY| clauses that expect an expression.

\color{black}
\subsection{Relational Algebra for Semantic Operators}

\begin{table*}[!t]
\centering
\footnotesize
\begin{tabular}{c l l l}
\toprule
\textbf{Operator} & \textbf{Definition} & \textbf{Output schema} & \textbf{Cardinality behavior} \\
\midrule

\fancycellProject{Project} &
$\pi^{s}_{P}(T) = \left\{ LLM(P, t[In]) \mid t \in T \right\}$ &
$schema(\pi^{s}_{P}(T)) = P.Out$ &
Preserved \\

\fancycellSelect{Select} &
$\sigma^{s}_{P}(T) = \left\{ t \in T \mid LLM(P, t[In]) = \text{True} \right\}$ &
$schema(\sigma^{s}_{P}(T)) = schema(T)$ &
Predicate selectivity \\

\fancycellRelation{Relation} &
$\rho^{s}_{P} = \left\{ t \in LLM(P) \right\}$ &
$schema(\rho^{s}_{P}) = P.Out$ &
Implicit predicate selectivity \\

\fancycellJoin{Join} &
$R \bowtie^{s}_{P} S = \left\{ (t_R, t_S) \mid LLM(P, t_R[In_R], t_S[In_S]) = \text{True} \right\}$ &
$schema(R \bowtie^{s}_{P} S) = schema(R) \cup schema(S)$ &
Join predicate selectivity \\

\fancycellAgg{Aggregate} &
$\gamma^{s}_{P}(T) = LLM(P, T[In])$ &
$schema(\gamma^{s}_{P}(T)) = P.Out$ &
Single value for table or group  \\

\bottomrule
\end{tabular}
\caption{Semantic operators and the corresponding relational algebra operators.}
\label{tab:semantic_operators}
\end{table*}

\dbname{} has an extended relational algebra operator for model inference that can interleave with regular relational operators. 
Let $f$ be a learned model that takes as input $x_1, \cdots, x_n$  and produces as output $y_1, \cdots, y_m$, i.e., 
$f(x_1, \cdots, x_n) = ({y_1, \cdots, y_m}).$
We define a relational algebra inference operator $
\delta_{y_1,...,y_m} (f, T_{x_1,...,x_n})$ that takes a user-defined learned model $f$ and the input attributes $x_1, \cdots, x_n$ of an input relation $T$ and produces the predicted attributes $y_1, \cdots, y_n$ by probing the learned model with the input attributes. 
Regardless of the model type, task, or the location in the query, it can be transformed into an equivalent $\delta$ operator in Relational Algebra.  $\delta$ is a minimal Relational Algebra operator for inference in a query.

For semantic queries that infer \verb|LLM|s, $\delta$  can be  extended further to semantic relational algebra operators, enabling logical transformations and equivalencies. For \verb|LLM|, the input is a well-formatted prompt embedded with input data values. Hence, we  define the prompt  $P = ~\langle instruction, In, Out \rangle$,  where $instruction$ is a textual instruction, e.g., a natural language condition, transformation, or generation request, $In$ are the prompt's input attributes, and $Out$ are the prompt's output attributes. Using the prompt, we  specify the structured input for the corresponding LLMs. Consider a tuple $t$ of Relation $T$ and a language model instance $f$. $
LLM(P, t) = f(instruction \oplus_{a \in In} t_{a})
$, 
where $\oplus$ is a string concatenation operation of input attributes, $a \in In$. Note that $schema(LLM(P, t)) = Out$.

Table~\ref{tab:semantic_operators} lists the semantic operators and their corresponding extended relational algebra, the output schema, and the cardinality of the operator. Semantic project projects the input attributes into the prompt's new semantic attributes. Semantic select filters tuples based on the prompt predicate and generates a new relation directly from the model, without an input relation. Semantic join joins two relations based on the prompt's semantic predicate. The semantic aggregate combines multiple input tuples into a single text output.

\subsection{Example Semantic Queries}
\label{subsec:usecases}
\input{3_usecases}

%% file: 3_usecases.tex
\lstset{style=SQLStyleTiny}

\begin{table*}[!t]
    \centering
    \footnotesize
    \begin{tabular}{l|l|c|l}
        \toprule
         & \textbf{Query} & \textbf{Operation} & \textbf{Type} \\
        \midrule
        Q1 & \begin{lstlisting}[language = SQL]
-- Extract the genre and main character of all Movies
SELECT title, genre, main_character
FROM LLM o4mini (PROMPT 'extract the {genre VARCHAR} and {main_character VARCHAR} from the {{plot}}', Movie);
\end{lstlisting} & \fancycellProject{Project} & Table \\
        \midrule
        Q2 & \begin{lstlisting}[language = SQL]
-- Find the language of all Movies
SELECT title, LLM o4mini (PROMPT 'what is the {language VARCHAR} of the movie {{title}}') FROM Movie;
\end{lstlisting} & \fancycellProject{Project} & Scalar \\
        \midrule
        Q3 & \begin{lstlisting}[language = SQL]
-- Generate table listing all maturity ratings the in US
CREATE TABLE MaturityRating AS SELECT maturity_label, description
FROM LLM o4mini (PROMPT 'Get all the maturity {maturity_label VARCHAR} and {description VARCHAR} in US');
\end{lstlisting} & \fancycellRelation{Relation} & Table \\
        \midrule
        Q4 & \begin{lstlisting}[language = SQL]
-- Find negative reviews about Titanic movie
SELECT r.review FROM Movie AS m NATURAL JOIN Review AS r
WHERE LLM o4mini (PROMPT 'is the sentiment of the {{r.review}} {negative BOOL}?') AND m.title = 'Titanic';
\end{lstlisting} & \fancycellSelect{Select} & Scalar \\
        \midrule
        Q5 & \begin{lstlisting}[language = SQL]
-- Find maturity rating of movies by understanding the plot
SELECT m.title, mr.maturity_label FROM Movie AS m JOIN MaturityRating AS mr
ON LLM o4mini (PROMPT 'is maturity rating {{mr.description}} depicted in the {{m.plot}}');
\end{lstlisting} & \fancycellJoin{Join} & Scalar \\
        \midrule
        Q6 & \begin{lstlisting}[language = SQL]
-- Summarize the cinematography style of each director
SELECT c.name, LLM AGG o4mini (PROMPT 'Summarize the cinematography {style VARCHAR} by the {{m.plot}}s')
FROM Cast AS c NATURAL JOIN Movie AS m
WHERE c.role = 'Director GROUP BY c.name;
\end{lstlisting} & \fancycellAgg{Aggregate} & Scalar \\
        \bottomrule
    \end{tabular}
    \caption{Example queries with the corresponding semantic operation and inference type}
    \label{tab:demo_table}
\end{table*}

\lstset{style=SQLStyle}

Table~\ref{tab:demo_table} lists six semantic SQL queries that depict different \dbname{} workloads. \emph{Q1} is a simple query with a semantic project to extract genre and main character of a movie from the movie plot. Semantic project is defined as a table inference. \emph{Q2} is a scalar counterpart of a semantic project, where the movie's language is inferred from the movie title. \emph{Q3} generates a new table by extracting knowledge from an LLM. \emph{Q4} has a semantic select to filter negative reviews of a specific movie based on sentiment analysis. \emph{Q5} joins two tables by an LLM-based semantic join condition. \emph{Q6} has a semantic aggregate to summarize each director's style. 
Observe that semantic operations coexist with traditional relational operators and adhere to the declarative nature of SQL.

%% file: 4_overview.tex
\begin{figure}[t]
    \centering
    \includegraphics[width=\linewidth]{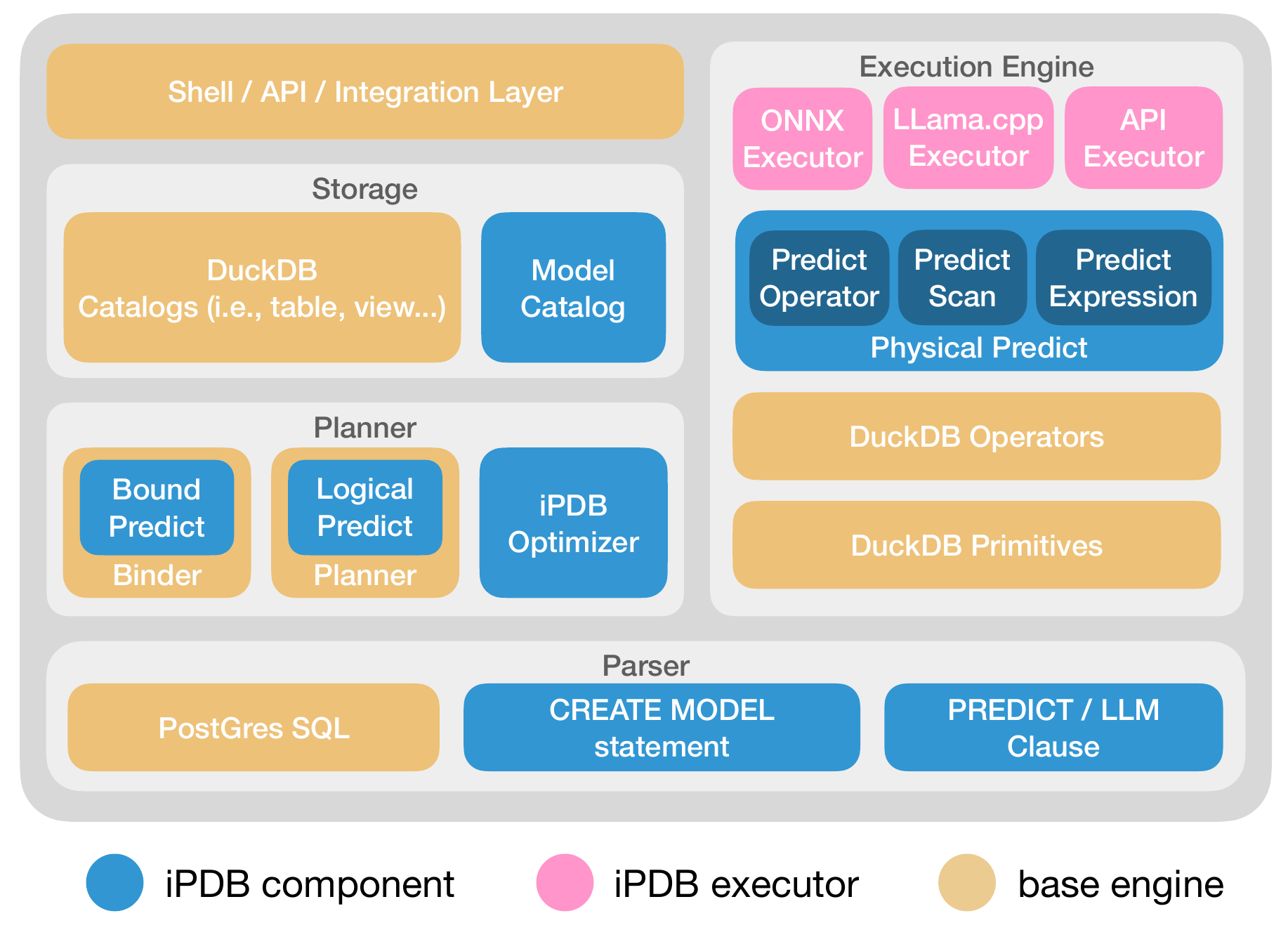}
    \vspace{-0.3in}
    \caption{\centering High-level architecture of \dbname{}}
    \label{fig:arch}
\end{figure}

\begin{figure*}[t]
    \centering
    \includegraphics[width=\linewidth]{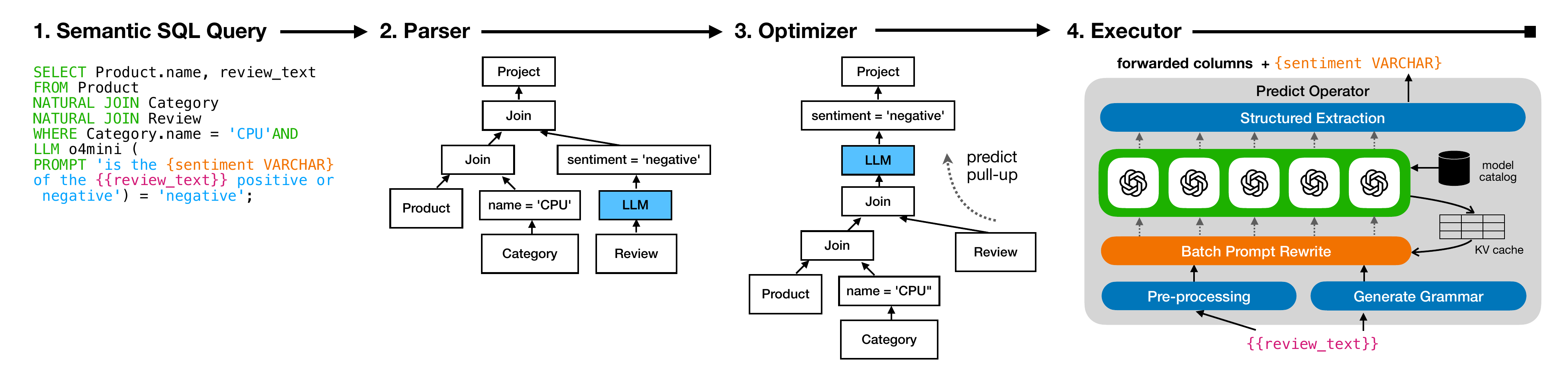}
    \caption{\centering Query Execution Process in \dbname{} to support semantic operators}
    \label{fig:flow}
\end{figure*}

\dbname{} is a new  system that supports semantic SQL and leverages DuckDB~\cite{duckdb_2019} as the base engine. We introduce the \verb|CREATE MODEL| statement, \verb|LLM|, and \verb|PREDICT| clauses to the parser as extensions to SQL. Native logical and physical LLM operators are implemented to execute semantic queries. DuckDB's system catalogs, table bindings, logical and physical planners, and  optimizer are extended to realize \dbname{}. 
Figures~\ref{fig:arch} and~\ref{fig:flow} give \dbname{'s} high-level components and query processing stages, respectively.
We choose DuckDB~\cite{duckdb_2019} as the foundation for \dbname{} for its analytical focus, lightweight design, and  execution architecture. DuckDB is a columnar, in-process analytical database optimized for analytical (OLAP) workloads, offering a vectorized execution engine over columnar data. This execution model aligns well with \dbname{’s} inference workloads.

\subsection{Model Catalog}

The model catalog (refer to Table~\ref{tab:model_cat}) is the central repository for storing metadata about \dbname{'s} all registered ML and LLMs. 

 \begin{table}[h]
    \centering
    \footnotesize
    \begin{tabular}{l|p{6.3cm}}
        \toprule
        \textbf{Attribute} & \textbf{Description} \\
        \midrule
        path & Model identifier for remote models or the path to the local model \\
    	type & Type of the model (i.e., LLM, TABULAR, EMBED) \\
	    on\_prompt & Flag if the model uses query time input/output resolution \\
    	base\_api & Base API address of remote model \\
    	secret & Secret name from the secrets catalog for proprietary models \\
	    relation & Name of the relation if the model is bound to a specific relation \\ 
	    input\_set & Input columns if inputs are defined at upload time \\
    	output\_set & Output columns if inputs are defined at upload time \\
        options & Key-value model specific options \\
        \bottomrule
    \end{tabular}
    \caption{Metadata stored in the Model Catalog.}
    \label{tab:model_cat}
\end{table}

The model catalog enables users to declaratively reference models at query time without having to configure the execution of inference models imperatively. The catalog helps ensure consistency and validation by verifying model availability, compatibility, and schema alignment at query time. This design provides the same convenience and reusability for models as first-class citizens, similar to traditional relational objects, e.g., tables and views. The model catalog is persisted by DuckDB's storage primitives.

\subsection{Semantic Query Parsing}

In \dbname{}, DuckDB's SQL parser is extended with new \verb|PREDICT|, and \verb|LLM| clauses. The parsed inference clauses are stored in a data structure that holds \verb|model_name|, \verb|source|, and \verb|prompt| to initialize an operator. \verb|model_name| is a qualified model that already exists in the model catalog. \verb|source| is a table reference if the input table is parsed (i.e., as an input relation for the \verb|PREDICT| clause or table inference \verb|LLM| clause). \verb|prompt| stores the prompt string if the \verb|LLM| clause is parsed.
During parsing, placeholders within prompts are resolved into explicit input (e.g., \verb|{{column}}|) and output (e.g., \verb|{column TYPE}|) attributes, enabling semantic operators to integrate seamlessly with typed relational columns. For models bound to a table or predefined feature, the input and output attributes are resolved from the model catalog.     
Upon parsing the query into an abstract syntax tree, the binding phase validates model reference against the catalog, and probes the base API URL, secret keys, and model-specific options. The input columns are checked against the bound columns by the descendant operators, and the output columns and types are informed to the ancestors to ensure compatibility with subsequent query operators. Thus, the model's properties are ready for the query optimizer phase.
\color{black}

\subsection{Semantic Query Planning}

Semantic queries are transformed into logical plans with {\tt PREDICT} operators alongside traditional relational operators. {\tt PREDICT} is realized by \verb|LogicalPredict| that extends DuckDB's logical operators. It stores the state required for inference: the model's type, name, path, API URL, secret, and execution options. Moreover, metadata for inference, e.g., names and types of input/output columns, are kept. 
Once the logical plan is materialized, DuckDB's optimizer is used to optimize traditional operations. However, without understanding that model inference is expensive, the traditional operator will attempt to reorder inference operators, believing they are zero-cost. \dbname{} has guardrails to inform the optimizer that {\tt PREDICT} is expensive and may prevent operator pushdown optimizations. \dbname{}  applies new logical optimizations for semantic operators that are discussed in more detail in Section~\ref{sec:optim}.
\color{black}

\subsection{Semantic Query Execution}

\dbname{} executes semantic queries via the physical operator {\tt PREDICT}.  {\tt PREDICT}'s behavior depends on the inference clause's position in the query. For table inference, it takes an input data chunk, processes the model, and populates the output data chunk. Scalar inference is realized by a {\tt PREDICT} expression executor to produce a vector instead of a data chunk, enabling it to integrate seamlessly with DuckDB's expression executor. A table generation initializes a {\tt PREDICT} scan to produce an output data chunk as a source.
For on-prompt queries, the prompt template and the passed input tuples are transformed into structured prompts that are sent to the target model. Model responses are parsed and validated against the expected schema. Predicted values are written to output data chunks to be passed to downstream operators. Error handling mechanisms can gracefully recover from malformed outputs, timeouts, or external service failures by retrying requests.
\color{black}

%% file: 5_phy_operator.tex
{\tt PREDICT} has three stages: the {\em configuration}, {\em loading}, and  {\em execution} stages. 
The {\em configuration} phase probes system defaults, session and model-specific options to configure {\tt PREDICT's} model execution. The {\em loading} phase either loads a locally stored model or initializes the API clients for remote models.  
In the {\em execution} stage, the inference operator adheres to push-based vectorized execution. {\tt PREDICT} executes once the sub-tree operators  aggregate a sufficient data chunk, and produces an output chunk that contains the processed dataset. {\tt PREDICT} is supplied with a consolidated data chunk from the input relations that consists of a set of rows organized as column vectors. It probes only the columns designated as inputs for the model. For LLM inference, it extracts the relevant input values and appends them to the prompt as key-value pairs.
{\tt PREDICT} pre-processes the prompt to ensure the generated output is optimal and structured (Section~\ref{subsec:struct_out}). 
The prompt is fed into the LLM to generate the desired results. {\tt PREDICT} extracts the structured output values from the LLM-generated results or probes the outputs of a model and populates them into the output data chunk.

\subsection{Structured Output Generation}\label{subsec:struct_out}

One challenge in incorporating LLMs into the relational model is obtaining consistent, schema-compliant output. \dbname{} augments user prompts with structural constraints, e.g., requiring outputs in a structured format to ensure that predictions are parsed reliably. Upon loading the physical predict operator, it first takes the prompt and output columns to determine an output format structure. 

\noindent
{\bf Grammar Forced Generation for Local Models:}
\dbname{} uses Context-Free grammar to guide and constrain sampling of local LLM models. LLMs' decoders use sampling parameters, e.g., \verb|top_p|, \verb|temperature|, to pick the next token from the top $k$ tokens. \dbname{} uses grammar-based sampling rules to constrain the top-k tokens. The grammar produces key typed value outputs. At execution, push-down automation samples the next tokens by filtering candidates that match one of the grammar rules.

{\bf Guided Generation for Remote Models:} Unlike local LLM models, \dbname{} cannot extend the generation algorithm of remote models. However, some models support a structured output format parameter in their API. Thus, the operator passes the output column names and types as a JSON schema, else, the operator falls back to guiding the LLM to produce structured outputs by instructing the model from system instructions to produce a parsable JSON object or array. It is instructed not to include any extra text, explanations, language specifiers, produce {<key>: <typed value>} pairs, and must be parsable by a standard parser.

{\bf Supported Data Types:}
To adhere to the concept of atomic attributes in relational databases, the inference should produce attributes with atomic values. It is trivial for tabular models (e.g., DNN models) that output a predefined set of typed columns. However, text-native models,
e.g., LLMs, produce text output regardless of the underlying type of the produced attributes. While any value returned by the LLM can be treated as text, this limits the operations applicable to the extracted attribute (e.g., a range predicate on an age value represented as a string). Thus, \dbname{} allows users to extract typed values from LLMs. \dbname{} instructs the model to produce parsable values for each type and post-processes the generated text to extract values in the corresponding type. Currently, \dbname{}  support s\verb|VARCHAR|, \verb|INTEGER|, \verb|DOUBLE|, and \verb|DATETIME|.

\subsection{Tuning and Configuring \dbname{}}

\dbname{} exposes configuration options to allow for tuning inference performance based on workload requirements. \dbname{} has \verb|batch size|, \verb|use batching|, and \verb|number of threads| as tunable options set by the \verb|SET <parameter> =| \verb|<value>;| command or specify parameters per model by providing key-value pairs in the \verb|OPTIONS| clause in the model upload command. Model-specific keyword arguments, e.g., temperature, maximum number of generated tokens, and top-p values, can be configured through model options at the upload time.

\subsection{Extensibility of \dbname{}}

\dbname{} supports a wide variety of tabular models through ONNX runtime~\cite{onnxruntime}, local LLM models through LLaMa.cpp, and remote LLM models through OpenAI-compatible APIs. Given the highly dynamic LLM model space, it is crucial to have an extensible platform  capable of adopting new model execution techniques. 

Due to the tight integration of \dbname{} with the base engine, the model catalog, native predict operator, and optimizer are implemented as internal system components.  
\dbname{} implements executors as statically linked third-party libraries. Each executor library extends a common interface that shares a common prediction state and a pre-defined set of functions that perform different phases of the {\tt PREDICT} operator. \dbname{} has a base \verb|Predictor| class including four member functions, namely {\tt Config($\cdot$)}, {\tt Load($\cdot)$}, {\tt PredictChunk($\cdot$)} and {\tt ScanChunk($\cdot$)}. To add a new execution technique, a developer extend the {\tt Predictor} class and implements and includes the new technique in the execution selection logic of the {\tt PREDICT} operator.

%% file: 6_optim.tex
\dbname{} aims to optimize semantic SQL queries to achieve minimal latency and cost within defined accuracy guarantees.

\subsection{Tuple Input Deduplication}
\label{subsec:deduplication}

Duplicate values exist in stored tables or can be produced during query execution. Invoking the LLM with the same prompt and inputs is redundant. \dbname{} utilizes a deduplication mechanism with a key-value cache in Operator  {\tt PREDICT}. {\tt PREDICT} maintains a concurrent hash table of input values and the LLM's parsed output throughout {\tt PREDICT}'s lifetime. Before calling the LLM, 
it checks if an existing cache value is hit for the given input values. This results in fewer LLM calls and fewer LLM tokens, hence reducing cost.

\subsection{{Multi-row Marshaling}}
\label{subsec:row_marshal}

Inferencing a model over a table needs to make a prediction for every tuple. This is slow and inefficient. 
If a set of tuples is batched to the model or prompted to the LLM in one call, it reduces the number of inference calls. 
\dbname{} takes advantage of DuckDB's vectorization to marshal a user-configurable number of tuple values to the model. An LLM is instructed to produce an array of predictions. \dbname{'s} multi-row prompt marshaling is enabled alongside prompt deduplication. If an input value of a tuple is already in the cache, the tuple is not included in the batch, but the cached LLM output is returned along with the other tuples in that batch.
\subsection{{Model Aware Input Ordering}}

This optimization reorders inputs before inference to better match the execution characteristics of the target model. For text-processing neural network models, \dbname{} sorts inputs by length so that shorter texts are grouped together in the same batch to reduce the size of the resulting feature matrices and improve inference efficiency. For LLMs, inputs are sorted in reverse order by length and are distributed over batches in round-robin strategy. This helps balance  workload over batches and parallel workers to prevent a batch from being dominated by unusually long prompts while others finish early. In both cases, the goal is to reduce inference overhead and improve overall throughput without changing query semantics.

\subsection{{Intra-Operator Parallelization}}
\dbname{} runs LLM calls in parallel over multiple threads based on the number of threads available system-wide or per model. This configuration enables parallelizing calls without exceeding the rate limits set by proprietary LLM vendors. However, during batched parallel execution, an entire batch may fail if one item in the batch fails. In this case, \dbname{} falls back to parallel execution of each tuple in the corresponding batch. Along with multi-row prompt marshaling and parallel LLM calls, \dbname{} reduces workload latency by a significant factor compared to serial per-tuple LLM execution.

\subsection{Semantic Select Pull-up}
In relational query optimization, the order in which predicates are applied is important. Having a highly selective predicate before expensive operations, e.g., sort or join, produces a more efficient query execution plan. Thus, traditional optimizers assume predicates as zero-latency operations and push them to the closest query sub-tree that contains the relation set of the predicate operation. This  optimization is invalid for semantic predicates due to their high execution costs. 
Thus, naively pushing LLM predicates down results in redundant LLM calls that may be removed by other traditional filters. Thus, \dbname{} pulls {\tt PREDICT} operators that contain a predicate on the predicted columns up the query tree. This results in fewer LLM calls that in turn reduces latency and cost.

\subsection{Semantic Select vs. Join Ordering}

A key distinction of semantic selection is that it can be selective yet expensive, unlike traditional filters that are virtually cost-free. Naively pushing semantic selects under  joins can result in redundant LLM calls as inference may be performed on tuples that are later discarded by the join. Whether the tuples get eliminated or not depends on which side of the join the semantic select is located (e.g., primary vs. foreign key side). \dbname{} optimizes both the semantic select and the joins by reordering semantic selects relative to  joins in a cost-aware manner, ensuring that expensive inference is only applied when necessary by pulling or pushing the semantic selects across a join. This either reduces redundant calls while preserving query semantics or reduces the join size. Section~\ref{subsec:selection_vs_join} analyzes the impact of ordering semantic selects vs. relational joins. 

\subsection{{{\tt LIMIT} Early Stopping}}

When a query has a LIMIT clause, \dbname{} can stop inference early once enough qualifying tuples have been produced. This is especially useful for semantic selects and semantic joins, where the cost of model inference dominates the rest of the plan. The base engine of \dbname{} already stops if enough tuples are produced. However, this is applied at a vector level, resulting in a full vector (i.e., 2048 tuples) execution before stopping that is excessive for semantic operators. \dbname{'s} optimizer propagates the LIMIT information to the \verb|PREDICT| operator, which in turn adopts smaller batch sizes (e.g., $2\times LIMIT$) and incrementally doubles until the maximum batch size is reached. This allows \verb|PREDICT| to halt prediction as soon as the output cardinality reaches the requested bound.

\subsection{{Cost-based Semantic Operator Ordering}}

Execution order of semantic operators within a query plan significantly influences performance. Inference latency is governed by model characteristics, instructions, and intermediate relation cardinality. A subpar ordering of semantic operators may exhibit significant latency disparities due to differing cost and selectivity. 

\dbname{} reorders semantic operators to reduce inference latency. The first stage applies existing heuristic transformations, e.g., predict pull-up and semantic select versus join ordering. The second stage considers adjacent semantic operators and applies either heuristic ordering between operator types or cost-based ordering within an operator type. Table~\ref{tbl:cost_mat} lists the heuristic and cost-based strategies applied across the various operator types.

\begin{table}[t]
\centering
\footnotesize
\begin{tabular}{l|cccccc}
\toprule
 & \textbf{Select} & \textbf{Join} & \textbf{Relation} & \textbf{Project} & \textbf{Agg.} \\
\midrule
\textbf{Select}    & Cost-based &  &  &  &  \\
\textbf{Join}      & Select & Cost-based &  &  &  \\
\textbf{Relation}  & Relation & Relation & N/A &  &  \\
\textbf{Project}   & Select & Join & Relation & Original &  \\
\textbf{Agg.} & Select & Join & Relation & Project & Original \\
\bottomrule
\end{tabular}
\caption{Semantic operator ordering strategy. Operator in a cell means operator is applied earlier. “Cost-based” refers to cost-based ordering decision. Ordering is not applicable ("N/A") if no two operators of such types can occur consecutively. “Original” keeps the existing order.}
\label{tbl:cost_mat}
\end{table}

In the cost-based strategy, \dbname{} evaluates a pair of consecutive semantic operators and prioritizes the semantic operator with the lower predicted cost. It uses linear regression trained to predict latency using features, including the number of instruction tokens, the number of output tokens (corresponding to the output type), and average input size on a sample of tuples. Additionally, for local models, we utilize the number of model parameters, context size, vocabulary size, number of layers, number of attention heads, hidden size, and quantization bits. Upon comparing each pair of consecutive semantic operators, the optimizer reorders the logic plan to apply the lower cost semantic operators first.

\subsection{{Semantic Join Decomposition}}

\begin{figure}[!ht]
\centering
\includegraphics[width=\linewidth]{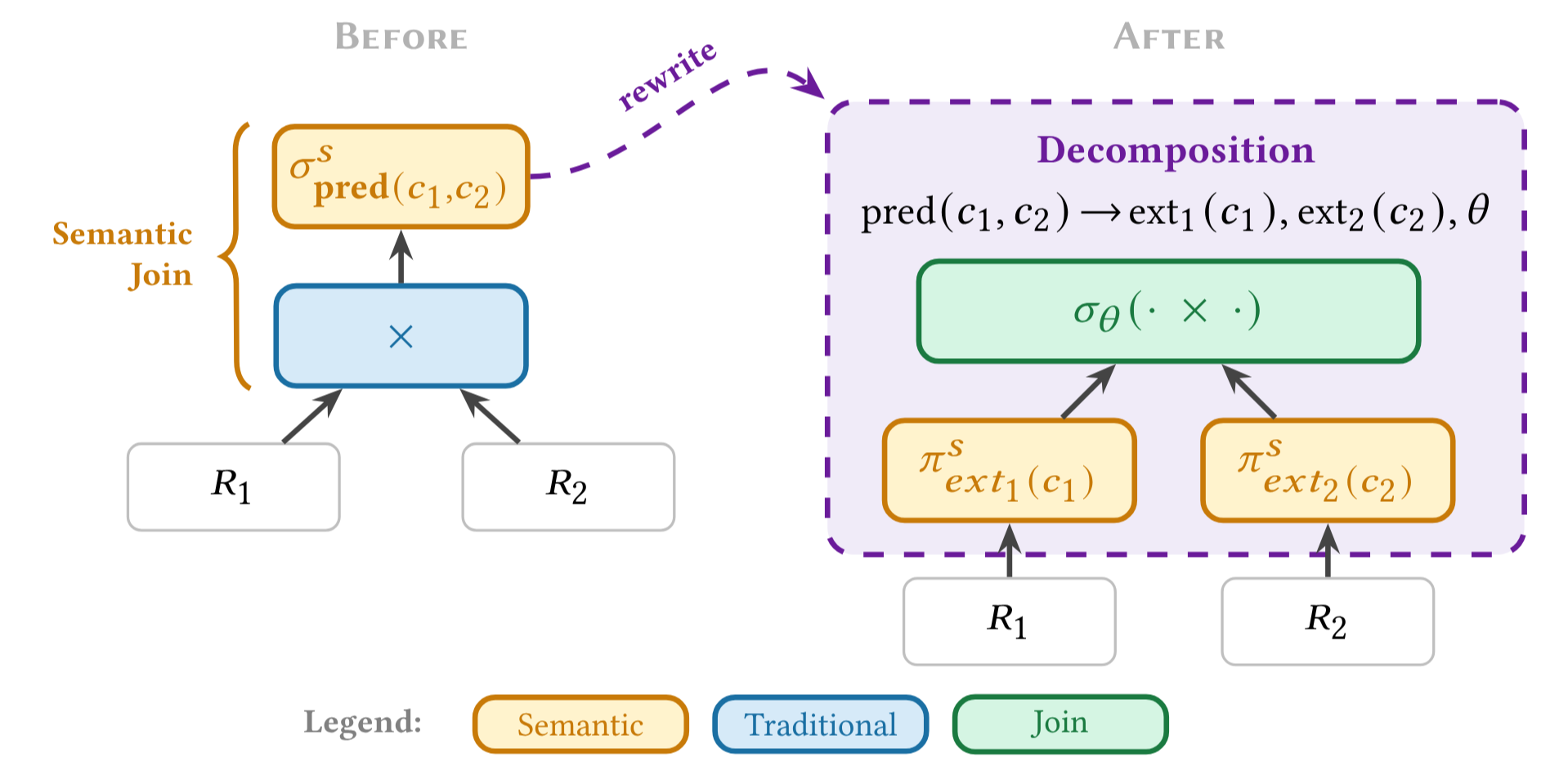}
\caption{Semantic Join Decomposition Mechanism.}
\label{fig:join_decomp_mech}
\end{figure}

Semantic joins represent a significant performance challenge as they require a cross product followed by an expensive semantic filter over all candidate tuple pairs. To mitigate this cost, inspired by ~\cite{decomp_join:2025}, \dbname{}'s optimizer attempts to decompose semantic joins into more efficient relational operations during query planning.

As in Figure~\ref{fig:join_decomp_mech}, the decomposition strategy uses the same LLM model to split the original semantic join predicate into two semantic projections followed by a traditional relational filter. The extracted predicate must preserve the semantic relationship expressed in the original prompt while enabling conventional relational join planning, e.g., a semantic join checking "CPU compatibility with Motherboard" may extract ‘socket type’ and ‘chipset family’ that can be matched deterministically. The optimizer ensures that input columns of the decomposed projections are available and output columns are compatible with the extracted predicate. If binding fails, \dbname{} falls back to the original semantic join plan. Thus,  \dbname{} opportunistically applies the optimization.
However, the approximation introduces a trade-off in quality and performance. The extracted features may not perfectly capture the original semantic relationship, potentially dropping some valid matches or introducing false positives. Users can toggle decomposition to prioritize performance vs. quality. The optimization is automatically disabled when a LIMIT clause is present as LIMIT's early stopping is more efficient on the original pairwise semantic filter than on the decomposed plan.

%% file: 8_evaluation.tex
We evaluate and compare \dbname{} against state-of-the-art systems across multiple datasets and several models. Furthermore, we evaluate the impact of optimizations introduced in Section~\ref{sec:optim}.

The experiments are run on an x86 Linux machine running Ubuntu 22.04.5 LTS distribution. The machine includes an Intel(R) Xeon(R) Platinum 8168 2.70GHz CPU and 2.95TB of memory. If the systems support parallel execution, the number of workers is set to 16. If the systems is capable of batching or row-marshaling, the size is set to 16. The local models are executed using GPU acceleration with an NVIDIA A30 with 24GB of memory.

{\bf Baselines.} We compare the following state-of-the-art systems:
\begin{itemize}
    \item[B1] \textbf{LOTUS}~\cite{lotus_2024} - has the maximum coverage of semantic operations on data frames. Lacks a declarative SQL interface and executes each operation eagerly.
    \item[B2] \textbf{Flock}~\cite{flock_2025} - LLM extension built for DuckDB that supports scalar and aggregate semantic functions.
    \item[B3] \textbf{BigQuery}~\cite{bigquery_2025} - BigQuery with AI functions enabled. 
    \item[B4] \textbf{\dbname{}} - Our optimized \dbname{} system.
\end{itemize}

{\bf Models.} We evaluate  \dbname{}  and other baselines on different models, different vendors, and execution environments.

\begin{itemize}
    \item[M1] \textbf{o4-mini} (Remote) - proprietary reasoning model by OpenAI~\cite{openai}.
    \item[M2] \textbf{Gemini 2.5 Flash} (Remote) - proprietary model by Google~\cite{gemini}.
    \item[M3] \textbf{Qwen 3.5 4B} (Local) - open-source model with 4B parameters that is executed locally.
\end{itemize}

{\bf Datasets.} We evaluate \dbname{} and the other baselines on the following datasets and benchmarks:
\begin{itemize}
    \item[D1] \textbf{PCParts} - The dataset of PC part inventory collected from PC Components~\cite{PCcomponents} and supplemented with product reviews in the SNAP Amazon Dataset~\cite{snap}. Consists of 5 tables with 2,060 total tuples.
    \item[D2] \textbf{SemanticMovies} - A benchmark with a subset of Join-Order Benchmark~\cite{jobs_2015} enriched with movie reviews that emphasize semantic workloads in the relational model. Consists of 8 tables with 842,593 total tuples.
    \item[D3] \textbf{SemBench}~\cite{lao2025sembench} - Benchmark proposed specifically for the evaluation of semantic operations. We evaluate the benchmark's 2000 scale version of the movie dataset.
\end{itemize}

\input{tables/benchmark_combined.tex}

\subsection{Performance on PCParts Dataset}

Table~\ref{tab:experimental_results_all} (D1) compares the performance of \dbname{} and the state-of-the-art on the PCParts queries. The latency, number of LLM calls, monetary cost, and the F1 score of the produced results are reported. Values indicated as \protect\fancyCellRed{n/a} are the workloads that are not supported by that particular system or model. If the cell is marked  \protect\fancycellYellowX{n/a}, the system consistently fails to produce an output, resulting in a runtime exception. Only the Gemini model results are reported for BigQuery (B3) as BigQuery does not use third party models. Additionally, the number of LLM calls made is not reported for BigQuery (B3) as it does not report this statistic.
Each system supports semantic operations except for semantic relation (Q2). 

\dbname{} outperforms the baseline systems consistently with regards to latency, while using less or comparable tokens, less number of LLM calls, and with comparable quality. LOTUS relies solely on parallelism to achieve low latency. In contrast, \dbname{} attempts to fit the optimal number of tuples into a single LLM call, and executes these calls in parallel. Thus, LOTUS invokes an LLM call per tuple in the input relation. This technique improves operator latency for relatively small tables. Additionally, it has to send metadata and formatting instructions with each call, resulting in a high total token count.
Flock optimizes LLM inference by concatenating input tuple values into batched LLM calls, resulting in fewer tokens and calls. However, Flock does not utilize logical optimizations and the returned results are not strictly structured, thus, resulting in higher latency and lower quality in some scenarios (e.g., Q3).
BigQuery performs with comparable quality to \dbname{} and has lower cost for semantic select workloads. However, the join decomposition optimization of \dbname{} results in $4x$ improvement on semantic join (Q4).
\dbname{'s} intra-operator and logical optimizations reduce the number of LLM calls and the number of tokens sent, resulting in low latency.
\color{black}

\subsection{Performance on SemanticMovies Dataset}

Table~\ref{tab:experimental_results_all} (D2) compares the performance of SemanticMovies queries. The benchmark contains seven relations that provide information on movies, related metadata, and reviews. Ground-truth labels are available for the movie's language and genre, and for the reviews' sentiment. Compared to the PCPart (D1) dataset, this dataset has 400x tuples, and requires inference at larger scale.
\dbname{} outperforms the baselines on latency, quality (F1 score) across the workloads while having comparable or better cost. Because LOTUS lacks a declarative SQL interface and logical optimizations, operators are executed in the optimal order through manual optimization. This dataset evaluates system's ability to process heavy semantic queries while being robust. Both LOTUS and Flock fail to produce consistent results for Q1 across models, where they throw an exception. The workload requires processing plotlines of movies, some of which contain graphical violence and mature content that the model refuses to process. When the model rejects a single tuple, the system fails the entire pipeline, whereas other systems handle the error gracefully. Additionally, across the workloads, LOTUS makes significantly more LLM calls and tokens due to the lack of row marshaling and deduplication.
In contrast, BigQuery is compatible with the workloads while having comparable quality. However, it has the largest latency due to the lack of logical optimizations, e.g., in Q3, BigQuery processes $\sim$91K tuples while others process only 320 by logical optimizations. Both baselines fail Q4 that contains a semantic relation operator.

\subsection{Performance on SemBench Benchmark}

Table~\ref{tab:experimental_results_all} (D3) compares the systems on the 2000 scale Movie dataset of the SemBench~\cite{lao2025sembench}. The latency, number of LLM calls, monetary cost, and the F1 score of the produced results are reported. 
Across the benchmark, \dbname{} is the most efficient system overall. It delivers the lowest or near-lowest query latencies, except for Q10, while avoiding the higher costs seen in the other systems. In contrast, LOTUS and Flock frequently incur higher runtime.
In Queries Q1, Q5, Q6, the workload has LIMIT clause requiring the systems to only produce sufficient number of tuples to fulfill the limit requirement. Both \dbname{} and BigQuery have optimizations to early stop on these scenarios. Hence, we observe significant difference in latency of \dbname{} and BigQuery vs. LOTUS and Flock. 
Queries Q2-Q4 and Q8-Q10 have traditional predicates that select tuples from the base table. Thus, all systems have the same number of tuples that are checked by the LLM and have comparatively similar execution plan. However, as visualized by the latency results, the intra-operator optimizations impact efficiency. LOTUS sends all the tuples to the LLM in parallel, resulting in higher costs. Flock batches the tuples into static batches and executes it in serial order. \dbname{} row marshals the tuples into dynamic batches. This increases the number of LLM calls, but it reduces the latency of a single parallel batch by distributing the tuples across workers.
Q5 is the most exhaustive query, requiring an unrestricted semantic join between $256 \times 256$ tuples. LOTUS and Flock timeout on Q5.  BigQuery applies a relatively efficient semantic join. \dbname{} utilizes  semantic join decomposition to split the join into a two extractions and a traditional join, which achieves $2\times$ performance while having a minor $0.043$ loss in quality.

\subsection{Comparison with AI-native Systems}
Several studies have implemented native semantic operator systems~\cite{palimpzestCIDR,docetl_2025}. These systems focus on document processing with only semantic operators. We compare \dbname{} against these systems on the BioDex~\cite{biodex_2023} benchmark that contains 65K biomedical articles. Expert annotated ground-truth is available from drug safety reports constructed from each article. The task is to classify the patient’s drug reactions given each medical article. This task is a large multi-label classification of 24K possible drug-reaction labels.

Table~\ref{tab:ai_native} shows, \dbname{} outperforms Palimpzest with higher rank-precision (RP) at 5, while producing with lower latency. Palimpzest have lower cost for using o4-mini; however, the difference is marginal. Both \dbname{} and Palimpzest consistently outperform DocETL. These results show that although \dbname{} is not designed for document workloads, it achieves comparable performance to the state of the art.

\begin{table}[!t]
    \centering
    \footnotesize
    \begin{tabular}{l|c|c|c|c}
        \toprule
        \textbf{System} & \textbf{Model} & \textbf{Cost} & \textbf{RP@5} & \textbf{Latency} \\
        \midrule
        \multirow{2}{*}{Palimpzest} & \MONE & \fancycellGreen{\$0.422} & \fancycellYellow{0.21619} & \fancycellYellow{452.15\,\text{s}}\\
         & \MTWO & \fancycellYellow{\$7.083} & \fancycellYellow{0.28293} & \fancycellYellow{811.96\,\text{s}}\\
        \midrule
        \multirow{2}{*}{DocETL} & \MONE & \fancycellYellow{\$2.094} & \fancycellYellow{0.16033}  & \fancycellYellow{1164.48\,\text{s}}\\
        & \MTWO & \fancycellYellow{\$9.560} & \fancycellYellow{0.22061}  & \fancycellYellow{2039.72\,\text{s}}\\
        \midrule
        \multirow{2}{*}{\dbname{}} & \MONE & \fancycellYellow{\$0.427} & \fancycellGreen{0.29554} & \fancycellGreen{371.74\,\text{s}}\\
        & \MTWO & \fancycellGreen{\$0.624} & \fancycellGreen{0.26536} & \fancycellGreen{743.24\,\text{s}} \\
        \bottomrule
    \end{tabular}
    \caption{Performance results on the BioDex dataset comparing the latency, monetary cost and rank-precision@5.}
    \label{tab:ai_native}
\end{table}

\subsection{Impact of Intra-Operator Optimizations}

\begin{figure}[t!]
    \centering
    
    \begin{subfigure}[b]{0.9\linewidth}
        \centering
        \includegraphics[width=\textwidth]{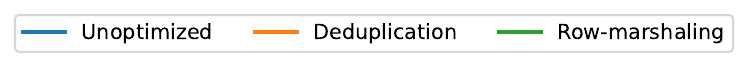}
    \end{subfigure}

    \begin{subfigure}[b]{0.48\linewidth}
        \centering
        \includegraphics[width=0.9\textwidth]{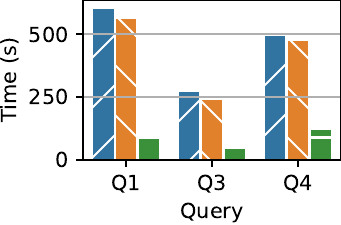}
        \caption{Latency (sequential)}
        \label{fig:plot1}
    \end{subfigure}
    \begin{subfigure}[b]{0.48\linewidth}
        \centering
        \includegraphics[width=0.9\textwidth]{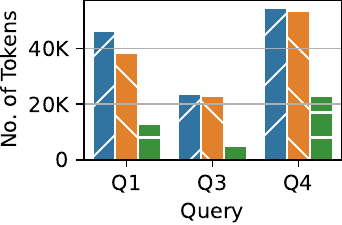}
        \caption{Total tokens (sequential)}
        \label{fig:plot2}
    \end{subfigure}
    
    \begin{subfigure}[b]{0.48\linewidth}
        \centering
        \includegraphics[width=0.9\textwidth]{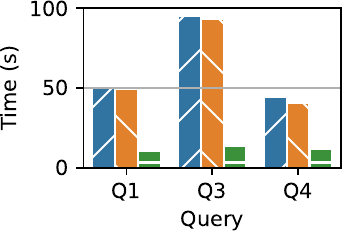}
        \caption{Latency (parallel)}
        \label{fig:plot1}
    \end{subfigure}
    \begin{subfigure}[b]{0.48\linewidth}
        \centering
        \includegraphics[width=0.9\textwidth]{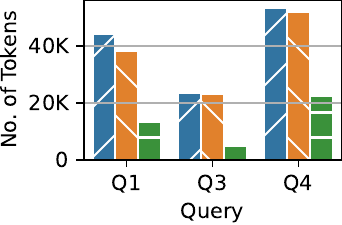}
        \caption{Total tokens (parallel)}
        \label{fig:plot2}
    \end{subfigure}

    \caption{Performance analysis of intra-operator optimizations under sequential and parallel (16 workers) execution.}
    \label{fig:intra_opt_res}
\end{figure}

In this experiment, we evaluate the impact of each intra-operator optimization. We control \dbname{} to allow only a single operator per run and report  operator latency and the total tokens used. We observe no difference in output quality when each optimization is turned on and off separately; thus, we do not report the accuracy values for this experiment. Additionally, we compare the impact of each optimization when it is run sequentially and in parallel. Figure~\ref{fig:intra_opt_res} gives the impact of each intra-operator optimization. For sequential experiments, Figure~\ref{fig:intra_opt_res}a gives the latency improvements against the unoptimized \dbname{}, and Figure~\ref{fig:intra_opt_res}b gives the total token usage improvement. Figures~\ref{fig:intra_opt_res}c and~\ref{fig:intra_opt_res}d present the respective improvements in the parallel setting. First, when comparing deduplication with the unoptimized operator, it is clear that the operator eliminates redundant LLM calls as evidenced by a reduction in total token usage across different queries. Consequently, operator latency  is reduced throughout different queries. We observe the same pattern in both sequential and parallel executions.
Secondly, when comparing row-marshaling with an unoptimized operator, row-marshaling achieves 5x improvement in token usage and latency on sequential executions. The token usage improvement comes from eliminating the redundant system and formatting instructions. Along with the lower number of input tokens, the row-marshaled operator has lower network overhead as it makes  fewer calls. Observe that even in the parallel setting, row-marshaling improves latency and token usage by 2x, highlighting the technique's advantage.

\subsection{Performance of Batched LLM API Calls}

\begin{figure}[t!]
    \centering
    \includegraphics[width=0.88\linewidth]{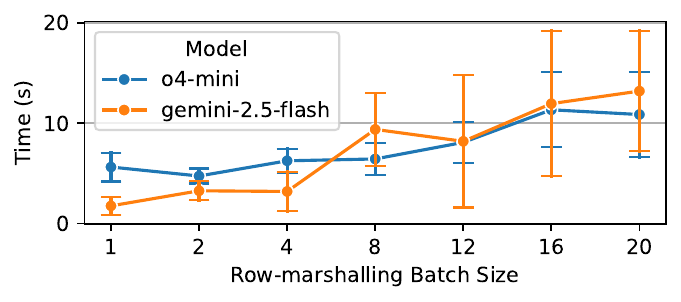}
    \caption{Generation latency vs row-marshaled batch size}
    \label{fig:row_marshal_latency}
\end{figure}

Figure~\ref {fig:row_marshal_latency} shows the tradeoff between the row-marshaling batch size, the number of rows concatenated into a single LLM prompt, and the inference latency for o4-mini and gemma-2.5-flash. The results indicate an increasing trend in latency as the batch size grows, consistent with the expectation that larger inputs result in a greater processing cost. This highlights a key trade-off in LLM-based query execution between input row-marshaling and responsiveness. While limited row-marshaling can improve throughput by reducing network overhead, excessive aggregation of rows can lead to latency growth and diminished performance consistency.

\subsection{Multi-row Batching vs Parallel LLM calls}

\begin{figure}[t]
    \centering
    \includegraphics[width=0.9\linewidth]{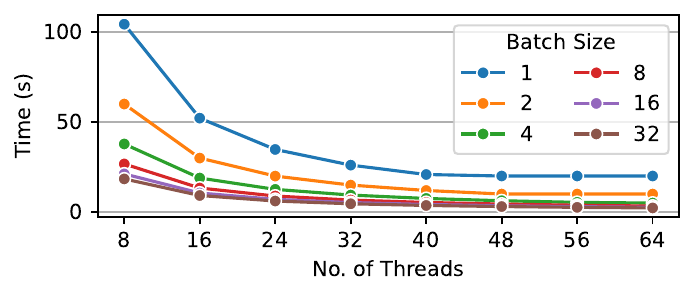}
    \caption{Query Latency of row-marshaling vs parallelization}
    \label{fig:row_marshal_vs_parallel}
\end{figure}

Figure~\ref{fig:row_marshal_vs_parallel} demonstrates the limitations of parallelization and how the limit can be overcome by row-marshaling even with the diminished performance of row-marshaled calls. The analysis assumes the number of tuples in the inference table is 10,000. We set the average request latency for different row-marshaling batch sizes based on the empirical results. Furthermore, the requests per minute limit is set to 500, which is the rate limit of the o4-mini model.

Observe that the blue line in the plot corresponds to the execution with disabled row-marshaling. The latency declines with the increasing parallelization until it stabilizes at 48 workers. The operation hits a limit due to the rate limit, regardless of increased parallelization. However, this limit can be overcome by row-marshaling as depicted by the other plot lines with increasing row-marshaling batch size. However, observe that improvement diminishes and is not linear to the batch size; this is a result of increased call latency for row-marshaled inputs. Thus, the operator will not benefit from overdoing the row concatenation. 

\subsection{Impact of {\tt PREDICT} Pulling}

\begin{figure}[t!]
    \centering
    \begin{subfigure}[b]{0.30\linewidth}
        \centering
        \includegraphics[width=\textwidth]{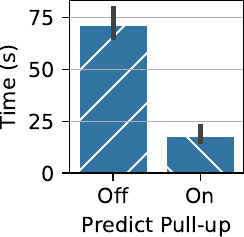}
    \end{subfigure}
    \begin{subfigure}[b]{0.315\linewidth}
        \centering
        \includegraphics[width=\textwidth]{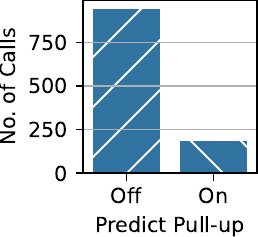}
    \end{subfigure}
    \begin{subfigure}[b]{0.33\linewidth}
        \centering
        \includegraphics[width=\textwidth]{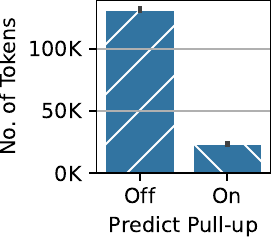}
    \end{subfigure}
    \caption{Performance  of {\tt PREDICT} pull-up optimization.}
    \label{fig:predict_pull}
\end{figure}

{\tt PREDICT} pulling is a heuristic logical optimization that aims to elevate expensive LLM-based filtering in the execution pipeline, assuming that traditional predicates and join operations will reduce the number of inference tuples. We evaluate the semantic select workload (i.e., D1:Q4) with and without {\tt PREDICT} pull-up. Figure~\ref{fig:predict_pull} gives the latency, number of calls, and number of total tokens used with and without {\tt PREDICT} pull up. Observe that the pull-up operation reduces the number of LLM calls. This is a result of the execution pipeline applying Predicate $category = `CPU`$  on  Products and the consequent join with Table Review, eliminating unrelated reviews. With fewer input tuples to the semantic operator, the overall latency and token usage are reduced.

\subsection{Impact of Semantic Select vs Join Ordering}
\label{subsec:selection_vs_join}

\begin{figure}[t!]

    \centering
    \begin{subfigure}[b]{0.45\linewidth}
        \includegraphics[width=\textwidth]{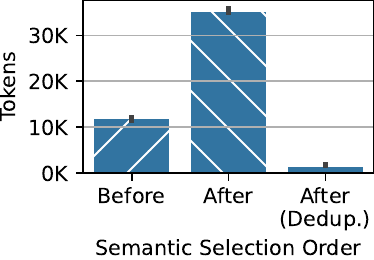}
    \end{subfigure}\hspace{2mm}
    \begin{subfigure}[b]{0.435\linewidth}
        \includegraphics[width=\textwidth]{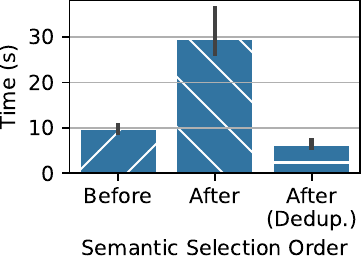}
    \end{subfigure}

    \caption{Impact of Semantic Select vs Join ordering.}
    \label{fig:sem_vs_join}
\end{figure}

In a one-to-many relationship, when a Semantic Select is on the foreign-key side, applying the Semantic Select before the join minimizes latency. Observe that the size of the resulting join is the cardinality of the foreign key relation. Since the LLM call is on the foreign-key side, the number of LLM calls is unaffected whether the join or the select is executed first. However, executing the select first reduces the join cost, resulting in an overall execution time of 49.81s versus 51.13s. Thus, the optimal strategy is to perform the select first.

If the select is on the primary-key side, as in Figure~\ref{fig:sem_vs_join}, then the semantic operator may be applied to duplicate entries on that side, since the resulting join has the cardinality of the foreign-key relation. The naive strategy would be to push select under the join to eliminate duplicate inference. However, the join condition in a one-to-many join may eliminate tuples on the primary-key side. Thus, using \dbname{'s} deduplication, the optimal strategy is to pull up the select and infer only the unique values of the primary key side.

In a many-to-many join, tuples under the semantic select can appear multiple times in the output of the join. A tuple may also not appear in the output relation. Thus, pushing select under the join can result in making redundant LLM calls. Thus, the best strategy is to pull up the select over the join. Deduplication eliminates the need to recompute outputs of multiple instances of the same input. 

\subsection{Cost-Aware Semantic Operator Ordering}

\begin{table}[h]
    \centering
    \footnotesize
    \begin{tabular}{l|c|c|c}
        \toprule
        \textbf{Order} & \textbf{Calls} & \textbf{Cost} & \textbf{Latency} \\
        \midrule
         $\sigma^s_{\text{'negative review'}} \rightarrow \sigma^s_{\text{'vendor is Intel'}}$ & \fancycellYellow{1067} & \fancycellYellow{\$0.169} & \fancycellYellow{25.61\,\text{s}}\\
         $\sigma^s_{\text{'vendor is Intel'}} \rightarrow \sigma^s_{\text{'negative review'}}$ & \fancycellGreen{350} & \fancycellGreen{\$0.034} & \fancycellGreen{9.59\,\text{s}}\\
        \bottomrule
    \end{tabular}
    \caption{Impact of ordering two Semantic Select operators.}
    \label{tbl:cost_workload}
\end{table}

Table~\ref{tbl:cost_workload} presents the performance gains from cost-based ordering of semantic operators. Executing the semantic selection to filter products by vendor results in the lower-cost operator being executed first while filtering the tuples for the relatively expensive review-sentiment-based filter. 

The three cost model approaches are evaluated by comparing the models' predicted orders with the ground truth. We use a dataset of 100 semantic operators, each with a different task, defined over the SemanticMovies (D2) dataset. For each random pair of operators, the ground truth is generated by calculating the end-to-end latency in the original and reversed order to identify the optimal order. The linear regression model on latency predicts the correct ordering with an accuracy of 0.764 and the average input tokens baseline has an accuracy of 0.712.

A key observation is that prediction errors are common when the two semantic operators differ only in marginal cost. In these cases, the operators' cost is insignificant for the optimal order, while selectivity has a stronger effect on the final end-to-end latency.

\subsection{Impact of {\tt LIMIT} early stopping}

\begin{figure}[t!]
    \centering
    \includegraphics[width=0.88\linewidth]{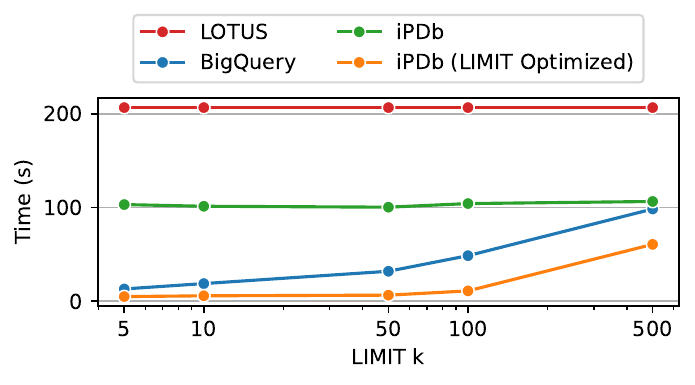}
    \caption{Latency vs varying LIMIT $k$ for SemBench (D3) Q1}
    \label{fig:limit_early_join}
\end{figure}

Figure~\ref{fig:limit_early_join} shows how  query latency changes as the LIMIT $k$ value increases. We use the SemBench (D3) Q1 query that already includes LIMIT 5, and we tune it for this evaluation. The results indicate that LOTUS is not well optimized for queries with LIMIT; this is due to eager execution of semantic operators, where the current system must fully execute the operator before taking k rows. Both BigQuery and \dbname{} optimize queries with LIMIT, and thus perform better for smaller $k$ values.
As $k$ grows, the runtime increases, with LIMIT Optimized, \dbname{} consistently provides the best performance. The unoptimized \dbname{} version is slower than its optimized counterpart, which highlights the benefit of LIMIT-aware execution.

\subsection{Impact of Semantic Join Decomposition}

\begin{table}[!t]
    \centering
    \footnotesize
    \setlength{\tabcolsep}{4pt}
    \begin{tabular}{l|ccc|ccc}
        \toprule
        \multirow{2}{*}{\textbf{Query}} & \multicolumn{3}{c|}{\textbf{without Decomposition}} & \multicolumn{3}{c}{\textbf{with Decomposition}} \\
         & \textbf{Cost} & \textbf{Quality} & \textbf{Latency} & \textbf{Cost} & \textbf{Quality} & \textbf{Latency} \\
        \midrule
        D1:Q5 & \fancycellYellow{\$0.19} & \fancycellGreen{0.560} & \fancycellYellow{141.03\,\text{s}} & \fancycellGreen{\$0.05} & \fancycellYellow{0.556} & \fancycellGreen{22.92\,\text{s}}\\
        D3:Q7 & \fancycellYellow{\$4.05} & \fancycellGreen{0.759} & \fancycellYellow{2539.06\,\text{s}} & \fancycellGreen{\$0.01} & \fancycellYellow{0.657} & \fancycellGreen{83.95\,\text{s}}\\
        \bottomrule
    \end{tabular}
    \caption{Semantic Join queries with and without Join Decomposition optimization}
    \label{tab:join_decomposition}
\end{table}

Table \ref{tab:join_decomposition} gives a comparison of semantic join queries executed with and without the Join Decomposition optimization. All experiments are conducted on \dbname{} using the o4-mini (\MONE) model. We report cost, result quality, and end-to-end latency.
The results show that enabling decomposition significantly reduces both cost and latency across all queries. For example, in D3:Q7, cost decreases from $0.05$ to $0.01$, and latency drops from 2539.06 seconds to 83.95 seconds. Similar improvements are observed for D1:Q5. While decomposition introduces a slight reduction in quality, the overall efficiency gains are substantial, demonstrating the effectiveness of semantic join decomposition. For workloads that require the best quality, this optimization can be disabled at the cost of higher latency. 

%% file: tables/benchmark_combined.tex
\newcommand{\MONE}{M1}
\newcommand{\MTWO}{M2}
\newcommand{\MTHREE}{M3}

\begin{table*}[htbp]
  \centering

  \begingroup
  \footnotesize
  \setlength{\tabcolsep}{2.7pt}
  \renewcommand{\arraystretch}{\arraystretchresulttable}
  \newcommand{\datasetCell}[1]{\multicolumn{14}{c}{\textbf{#1}}}

  \begin{tabular*}{\textwidth}{l|c|@{\extracolsep{\fill}}
      c@{\hspace{0.1em}}c@{\hspace{0.1em}}c@{\hspace{1em}}|
      c@{\hspace{0.1em}}c@{\hspace{0.1em}}c@{\hspace{1em}}|
      c@{\hspace{0.1em}}c@{\hspace{0.1em}}c@{\hspace{1em}}|
      c@{\hspace{0.1em}}c@{\hspace{0.1em}}c}

    \EvalHeader

    \datasetCell{(D1) PC Parts} \\
    \midrule
    
     \multirow{3}{*}{\fancycellProject{Q1}} & \MONE & \fancycellYellow{208|\$0.402} & \fancycellYellow{0.490} & \fancycellYellow{67.50\,\text{s}} &  \fancycellYellow{13|\$0.046} & \fancycellGreen{1.000} & \fancycellYellow{59.46\,\text{s}} & \fancyCellRed{n|a} & \fancyCellRed{n|a} & \fancyCellRed{n|a} & \fancycellGreen{16|\$0.044} & \fancycellGreen{1.000} & \fancycellGreen{9.30\,\text{s}} \\
     & \MTWO & \fancycellYellow{208|\$0.014} & \fancycellYellow{0.423} & \fancycellYellow{24.37\,\text{s}} & \fancycellYellow{13|\$0.006} & \fancycellGreen{1.000} & \fancycellYellow{29.91\,\text{s}} & \fancycellGreen{-|\$0.012} & \fancycellGreen{1.000} & \fancycellYellow{15.10s} & \fancycellYellow{16|\$0.005} & \fancycellGreen{1.000} & \fancycellGreen{6.47\,\text{s}} \\
     & \MTHREE & \fancycellYellow{208|\$0.000} & \fancycellYellow{0.856} & \fancycellYellow{32.25\,\text{s}} & \fancycellGreen{13|\$0.000} & \fancycellYellow{0.900} & \fancycellYellow{54.29\,\text{s}} & \fancyCellRed{n|a} & \fancyCellRed{n|a} & \fancyCellRed{n|a} & \fancycellYellow{16|\$0.000} & \fancycellGreen{0.995} & \fancycellGreen{5.56\,\text{s}} \\
    \midrule
    \multirow{3}{*}{\fancycellRelation{Q2}} & \MONE & \fancyCellRed{n|a} & \fancyCellRed{n|a} & \fancyCellRed{n|a} & \fancyCellRed{n|a} & \fancyCellRed{n|a} & \fancyCellRed{n|a} & \fancyCellRed{n|a} & \fancyCellRed{n|a} & \fancyCellRed{n|a} & \fancycellGreen{1|\$0.009} & \fancycellGreen{1.000} & \fancycellGreen{16.18\,\text{s}} \\
    & \MTWO & \fancyCellRed{n|a} & \fancyCellRed{n|a} & \fancyCellRed{n|a} & \fancyCellRed{n|a} & \fancyCellRed{n|a} & \fancyCellRed{n|a} & \fancyCellRed{n|a} & \fancyCellRed{n|a} & \fancyCellRed{n|a} & \fancycellGreen{1|\$0.003} & \fancycellGreen{1.000} & \fancycellGreen{7.29\,\text{s}} \\
     & \MTHREE & \fancyCellRed{n|a} & \fancyCellRed{n|a} & \fancyCellRed{n|a} & \fancyCellRed{n|a} & \fancyCellRed{n|a} & \fancyCellRed{n|a} & \fancyCellRed{n|a} & \fancyCellRed{n|a} & \fancyCellRed{n|a} & \fancycellGreen{1|\$0.000} & \fancycellGreen{1.000} & \fancycellGreen{0.16\,\text{s}} \\
    \midrule
    \multirow{3}{*}{\fancycellProject{Q3}} & \MONE & \fancycellYellow{208|\$0.421} & \fancycellYellow{0.000} & \fancycellYellow{79.01\,\text{s}} & \fancycellGreen{3|\$0.011} & \fancycellYellow{0.512} & \fancycellYellow{16.37\,\text{s}} &  \fancyCellRed{n|a} & \fancyCellRed{n|a} & \fancyCellRed{n|a} & \fancycellGreen{3|\$0.043} & \fancycellGreen{0.951} & \fancycellGreen{15.98\,\text{s}} \\
    & \MTWO & \fancycellYellow{208|\$0.051} & \fancycellYellow{0.010} & \fancycellYellow{47.09\,\text{s}} & \fancycellYellow{3|\$0.002} & \fancycellYellow{0.121} & \fancycellYellow{34.83\,\text{s}} & \fancycellGreen{-|\$0.001} & \fancycellYellow{0.829} & \fancycellYellow{22.35\,\text{s}} & \fancycellYellow{3|\$0.002} & \fancycellGreen{0.890} & \fancycellGreen{13.11\,\text{s}} \\
     & \MTHREE & \fancycellYellow{208|\$0.000} & \fancycellYellow{0.038} & \fancycellYellow{23.79\,\text{s}} & \fancycellGreen{3|$0.000$} & \fancycellYellow{0.164} & \fancycellYellow{19.77\,\text{s}} & \fancyCellRed{n|a} & \fancyCellRed{n|a} & \fancyCellRed{n|a} & \fancycellGreen{3|\$0.000} & \fancycellGreen{0.536} & \fancycellGreen{1.40\,\text{s}} \\
    \midrule
    \multirow{3}{*}{\fancycellSelect{Q4}} & \MONE & \fancycellYellow{192|\$0.131} & \fancycellYellow{0.963} & \fancycellYellow{34.74\,\text{s}} & \fancycellYellow{42|\$0.190} & \fancycellYellow{0.970} & \fancycellYellow{293.29\,\text{s}} &  \fancyCellRed{n|a} & \fancyCellRed{n|a} & \fancyCellRed{n|a} & \fancycellGreen{16|\$0.073} & \fancycellGreen{0.996} & \fancycellGreen{14.52\,\text{s}} \\
    & \MTWO & \fancycellYellow{192|\$0.035} & \fancycellYellow{0.963} & \fancycellYellow{41.25\,\text{s}} & \fancycellYellow{42|\$0.026} & \fancycellYellow{0.960} & \fancycellYellow{330.03\,\text{s}}  & \fancycellYellow{-|\$0.037} & \fancycellYellow{0.948} & \fancycellYellow{11.48\,\text{s}} & \fancycellGreen{16|\$0.007} & \fancycellGreen{0.968} & \fancycellGreen{8.45\,\text{s}} \\
     & \MTHREE & \fancycellYellow{192|\$0.000} & \fancycellYellow{0.162} & \fancycellYellow{96.68\,\text{s}} & \fancycellYellow{42|\$0.000} & \fancycellYellow{0.664} & \fancycellYellow{175.28\,\text{s}} & \fancyCellRed{n|a} & \fancyCellRed{n|a} & \fancyCellRed{n|a} & \fancycellGreen{16|\$0.000} & \fancycellGreen{0.927} & \fancycellGreen{4.28\,\text{s}} \\
    \midrule
    \multirow{3}{*}{\fancycellJoin{Q5}} & \MONE & \fancycellYellow{1.6K|\$2.157} & \fancycellYellow{0.498} & \fancycellYellow{464.03\,\text{s}} & \fancycellYellow{123|\$0.835} & \fancycellGreen{0.572} & \fancycellYellow{1046.25\,\text{s}} & \fancyCellRed{n|a} & \fancyCellRed{n|a} & \fancyCellRed{n|a} & \fancycellGreen{14|\$0.050} & \fancycellYellow{0.556} & \fancycellGreen{22.91\,\text{s}} \\
    & \MTWO & \fancycellYellow{1.6K|\$0.339} & \fancycellYellow{0.575} & \fancycellYellow{260.97\,\text{s}} & \fancycellYellow{123|\$0.050} & \fancycellYellow{0.501} & \fancycellYellow{745.55\,\text{s}} & \fancycellYellow{-|\$0.040} & \fancycellYellow{0.574} & \fancycellYellow{101.25} & \fancycellGreen{14|\$0.003} & \fancycellGreen{0.591} & \fancycellGreen{25.99\,\text{s}} \\
     & \MTHREE & \fancycellYellow{1.6K|\$0.000} & \fancycellYellow{0.332} & \fancycellYellow{194.56\,\text{s}} & \fancycellYellow{123|\$0.000} & \fancycellYellow{0.448} & \fancycellYellow{626.76\,\text{s}} & \fancyCellRed{n|a} & \fancyCellRed{n|a} & \fancyCellRed{n|a} & \fancycellGreen{106|\$0.000} & \fancycellGreen{0.602} & \fancycellGreen{67.28\,\text{s}} \\
    \midrule

    \datasetCell{(D2) SemanticMovies} \\
    \midrule
    
      \multirow{3}{*}{\fancycellProject{Q1}} & \MONE & \fancycellYellow{6.7K|\xmark} & \fancycellYellowX{\xmark} & \fancycellYellow{1874.32\,\text{s}} & \fancycellYellow{420|\xmark} & \fancycellYellowX{n/a} & \fancycellYellow{14.26\,\text{s}} & \fancyCellRed{n|a} & \fancyCellRed{n|a} & \fancyCellRed{n|a} & \fancycellGreen{514|\$2.167} & \fancycellGreen{0.354} & \fancycellGreen{798.04\,\text{s}} \\
     & \MTWO & \fancycellYellow{6.7K|\xmark} & \fancycellYellowX{\xmark} & \fancycellYellow{1260.45\,\text{s}} & \fancycellYellow{420|\xmark} & \fancycellYellowX{n/a} & \fancycellYellow{16.82\,\text{s}} & \fancycellYellow{-|\$44.946} & \fancycellYellow{0.391} & \fancycellYellow{1295.41\,\text{s}} & \fancycellGreen{514|\$1.224} & \fancycellGreen{0.563} & \fancycellGreen{443.62\,\text{s}} \\
     & \MTHREE & \fancycellYellow{6.7K|\$0.000} & \fancycellYellow{0.446} & \fancycellYellow{808.02\,\text{s}} & \fancycellYellow{420|$0.000$} & \fancycellYellow{0.250} & \fancycellYellow{2424.02\,\text{s}} & \fancyCellRed{n/a} & \fancyCellRed{n/a} & \fancyCellRed{n/a} & \fancycellGreen{514|\$1.224} & \fancycellGreen{0.563} & \fancycellGreen{443.62\,\text{s}} \\
    \midrule
      \multirow{3}{*}{\fancycellProject{Q2}} & \MONE & \fancycellYellow{5K|\xmark} & \fancycellYellowX{\xmark} & \fancycellYellow{2153.14\,\text{s}} & \fancycellGreen{313|\$0.873} & \fancycellYellow{0.896} & \fancycellYellow{1305.30\,\text{s}} & \fancyCellRed{n|a} & \fancyCellRed{n|a} & \fancyCellRed{n|a} & \fancycellYellow{331|\$1.092} & \fancycellGreen{0.950} & \fancycellGreen{347.10\,\text{s}} \\
     & \MTWO & \fancycellYellow{5K|\$1.056} & \fancycellYellow{0.305} & \fancycellYellow{656.13\,\text{s}} & \fancycellYellow{313|\$0.255} & \fancycellYellow{0.848} & \fancycellYellow{2480.07\,\text{s}} & \fancycellYellow{-|\$25.306} & \fancycellYellow{0.895} & \fancycellYellow{925.03\,\text{s}} & \fancycellGreen{331|\$0.227} & \fancycellGreen{0.954} & \fancycellGreen{129.23\,\text{s}} \\
     & \MTHREE & \fancycellYellow{5K|\$0.000} & \fancycellYellow{0.671} & \fancycellYellow{600.42\,\text{s}} & \fancycellGreen{331|\$0.000} & \fancycellYellow{0.560} & \fancycellYellow{1873.31\,\text{s}} & \fancyCellRed{n/a} & \fancyCellRed{n/a} & \fancyCellRed{n/a} & \fancycellGreen{331|\$0.000} & \fancycellGreen{0.779} & \fancycellGreen{152.97\,\text{s}} \\
    \midrule
      \multirow{3}{*}{\fancycellSelect{Q3}} & \MONE &  \fancycellYellow{320|\$0.089} & \fancycellYellow{0.852} & \fancycellYellow{85.27\,\text{s}} & \multicolumn{3}{c|}{\fancycellGray{Timeout > 5\,\text{h}}} & \fancyCellRed{n|a} & \fancyCellRed{n|a} & \fancyCellRed{n|a} & \fancycellGreen{20|\$0.074} & \fancycellGreen{0.922} & \fancycellGreen{24.01\,\text{s}} \\
     & \MTWO & \fancycellYellow{320|\$0.129} & \fancycellYellow{0.967} & \fancycellYellow{62.76\,\text{s}} & \multicolumn{3}{c|}{\fancycellGray{Timeout > 5\,\text{h}}} & \fancycellYellow{-|\$92.558} & \fancycellYellow{0.955} & \fancycellYellow{15,540.35\,\text{s}} & \fancycellGreen{20|\$0.047} & \fancycellGreen{0.969} & \fancycellGreen{17.25\,\text{s}} \\
      & \MTHREE & \fancycellYellow{320|\$0.000} & \fancycellYellow{0.588} & \fancycellYellow{159.62\,\text{s}} & \multicolumn{3}{c|}{\fancycellGray{Timeout > 5\,\text{h}}} & \fancyCellRed{n/a} & \fancyCellRed{n/a} & \fancyCellRed{n/a} & \fancycellGreen{20|\$0.000} & \fancycellGreen{0.882} & \fancycellGreen{5.48\,\text{s}} \\
    \midrule
      \multirow{3}{*}{\fancycellRelation{Q4}} & \MONE & \fancyCellRed{n|a} & \fancyCellRed{n|a} & \fancyCellRed{n|a} & \fancyCellRed{n|a} & \fancyCellRed{n|a} & \fancyCellRed{n|a} & \fancyCellRed{n|a} & \fancyCellRed{n|a} & \fancyCellRed{n|a} & \fancycellGreen{1|\$0.003} & \fancycellGreen{1.000} & \fancycellGreen{4.91\,\text{s}} \\
     & \MTWO & \fancyCellRed{n|a} & \fancyCellRed{n|a} & \fancyCellRed{n|a} & \fancyCellRed{n|a} & \fancyCellRed{n|a} & \fancyCellRed{n|a} & \fancyCellRed{n|a} & \fancyCellRed{n|a} & \fancyCellRed{n|a} & \fancycellGreen{1|\$0.001} & \fancycellGreen{1.000} & \fancycellGreen{2.65\,\text{s}} \\
     & \MTHREE & \fancyCellRed{n|a} & \fancyCellRed{n|a} & \fancyCellRed{n|a} & \fancyCellRed{n|a} & \fancyCellRed{n|a} & \fancyCellRed{n|a} & \fancyCellRed{n|a} & \fancyCellRed{n|a} & \fancyCellRed{n|a} & \fancycellGreen{1|\$0.000} & \fancycellGreen{1.000} & \fancycellGreen{2.01\,\text{s}} \\
    \midrule

    \datasetCell{(D3) SemBench (Movie)} \\
    \midrule

    \fancycellSelect{Q1} & \MTWO & \fancycellYellow{2K|\$1.510} & \fancycellGreen{1.000} & \fancycellYellow{206.67\,\text{s}} &  \fancycellYellow{125|\$0.058} & \fancycellGreen{1.000} & \fancycellYellow{880.18\,\text{s}} & \fancycellYellow{
-|\$0.050} & \fancycellGreen{1.000} & \fancycellYellow{26.30\,\text{s}} & \fancycellGreen{10|\$0.001} & \fancycellGreen{1.000} & \fancycellGreen{3.74\,\text{s}} \\
        
    \fancycellSelect{Q2} & \MTWO &\fancycellYellow{120|\$0.080} & \fancycellGreen{1.000} & \fancycellYellow{14.4\,\text{s}} & \fancycellYellow{8|\$0.030} & \fancycellYellow{0.526} & \fancycellYellow{47.64\,\text{s}} & \fancycellGreen{-|\$0.003} & \fancycellGreen{1.000} & \fancycellYellow{9.50\,\text{s}} & \fancycellYellow{15|\$0.004} & \fancycellGreen{1.000} & \fancycellGreen{6.70\,\text{s}} \\
    
    \fancycellSelect{Q3} & \MTWO & \fancycellYellow{120|\$0.090} & \fancycellYellow{0.640} & \fancycellYellow{14.8\,\text{s}} & \fancycellYellow{8|\$0.030} & \fancycellYellow{0.429} & \fancycellYellow{42.28\,\text{s}} & \fancycellGreen{-|\$0.003} & \fancycellYellow{0.640} & \fancycellYellow{11.00\,\text{s}} & \fancycellYellow{15|\$0.004} & \fancycellGreen{0.833} & \fancycellGreen{8.67\,\text{s}}  \\
    
    \fancycellSelect{Q4} & \MTWO &\fancycellYellow{120|\$0.070} & \fancycellYellow{0.640} & \fancycellYellow{13.0\,\text{s}} & \fancycellYellow{8|\$0.040} & \fancycellYellow{0.715} & \fancycellYellow{44.34\,\text{s}} & \fancycellGreen{-|\$0.003} & \fancycellYellow{0.640} & \fancycellYellow{11.40\,\text{s}} & \fancycellGreen{15|\$0.003} & \fancycellGreen{0.738} & \fancycellGreen{6.43\,\text{s}} \\

    \fancycellJoin{Q5} & \MTWO & \fancycellYellow{25K|\$32.151} & \fancycellYellow{0.666} & \fancycellYellow{3950.92\,\text{s}} & \fancycellYellow{127|\$0.101} & \fancycellGreen{1.000} & \fancycellYellow{1878.42\,\text{s}} & \fancycellYellow{-|\$1.010} & \fancycellYellow{0.890} & \fancycellYellow{54.50\,\text{s}} & \fancycellGreen{10|\$0.001} & \fancycellGreen{1.000} & \fancycellGreen{7.28\,\text{s}} \\
    
    \fancycellJoin{Q6} & \MTWO &\fancycellYellow{25K|\$32.634} & \fancycellYellow{0.650} & \fancycellYellow{4286.42\,\text{s}} & \fancycellYellow{127|\$0.102} & \fancycellYellow{0.697} & \fancycellYellow{1861.59\,\text{s}} & \fancycellYellow{-|\$1.000} & \fancycellYellow{0.690} & \fancycellYellow{54.50\,\text{s}} & \fancycellGreen{19|\$0.003} & \fancycellGreen{0.793} & \fancycellGreen{23.88\,\text{s}}\\
    
    \fancycellJoin{Q7} & \MTWO & \fancycellYellow{32K|\$39.873} & \fancycellYellow{0.231} & \fancycellYellow{5084.03\,\text{s}} & \multicolumn{3}{c|}{\fancycellGray{Timeout > 2\,\text{h}}} & \fancycellYellow{-|\$3.310} & \fancycellGreen{0.700} & \fancycellYellow{198.30\,\text{s}}  & \fancycellGreen{29|\$0.007} & \fancycellYellow{0.657} & \fancycellGreen{83.95\,\text{s}}\\

    \fancycellProject{Q8} & \MTWO & \fancycellYellow{120|\$0.050} & \fancycellYellow{0.930} & \fancycellYellow{14.3\,\text{s}} & \fancycellYellow{8|\$0.030} & \fancycellGreen{0.833} & \fancycellYellow{50.69\,\text{s}} & \fancycellGreen{-|\$0.003} & \fancycellYellow{0.760} & \fancycellYellow{10.90\,\text{s}} & \fancycellGreen{15|\$0.003} & \fancycellYellow{0.781} & \fancycellGreen{8.26\,\text{s}}\\
    
    \fancycellProject{Q9} & \MTWO & \fancycellYellow{260|\$0.030} & \fancycellYellow{0.750} & \fancycellYellow{49.44\,\text{s}} &  \fancyCellRed{n|a} & \fancyCellRed{n|a} & \fancyCellRed{n|a} & \fancycellYellow{-|\$0.020} & \fancycellYellow{0.780} & \fancycellYellow{13.30\,\text{s}} & \fancycellGreen{16|\$0.005} & \fancycellGreen{1.000} & \fancycellGreen{12.57\,\text{s}} \\
    
    \fancycellProject{Q10} & \MTWO & \fancycellYellow{2K|\$2.200} & \fancycellYellow{0.400} & \fancycellYellow{284.9\,\text{s}} &  \fancyCellRed{n|a} & \fancyCellRed{n|a} & \fancyCellRed{n|a} & \fancycellGreen{-|\$0.130} & \fancycellYellow{0.440} & \fancycellGreen{32.10\,\text{s}} & \fancycellGreen{104|\$0.066} & \fancycellGreen{0.525} & \fancycellYellow{42.58\,\text{s}} \\
    \bottomrule

  \end{tabular*}
  \endgroup
  \caption{Performance, cost and quality of \dbname{} compared to state-of-the-art systems on varying benchmarks. Operator Type: \protect\fancycellSelect{Semantic Select}, \protect\fancycellProject{Semantic Project}, \protect\fancycellRelation{Semantic Relation}, and \protect\fancycellJoin{Semantic Join}. Labels: \protect\fancycellGreen{Best}, \protect\fancycellYellow{Regular}, \protect\fancycellYellowX{n/a} if consistent runtime failures and \protect\fancyCellRed{Not supported} for not supported. Models: \MONE - o4-mini, \MTWO - Gemini-2.5-flash, M3 - Qwen 3.5 4B (local).}
  \label{tab:experimental_results_all}
\end{table*}

%% file: 9_related_work.tex
Several recent studies aim to address the execution of ML and semantic queries in databases.

{\bf Document Processing Systems: }
Palimpzest~\cite{palimpzestCIDR} implements a physical optimization scheme for semantic operators for document processing. The user defines the imperative set of semantic operators to analyze a document, which will be mapped to a logical plan. The key contribution in the work is the physical optimizer that selects the best model for each semantic operator for a given optimization policy. DocETL~\cite{docetl_2025} offers a configuration interface where users can plan for the semantic operations. These systems have shown promising results in document processing, but they are not ideal for structured data in the relational model. They do not provide a declarative SQL-like interface, and queries must be defined imperatively. Both systems only allow semantic operations, lacking the ability to utilize traditional operations.

{\bf Systems with ML/LLM Augmentations:}
Several research and production-ready systems have introduced LLM augmented user-defined functions. This enables users to conveniently write model inference logic in Python and use the UDF interface to bind it with the database. However, in a generic UDF implementation, the database treats the inference function as a black box that limits its optimization capabilities. Additionally, the user has to implement and optimize how the model is inferred in the UDFs. Several recent works have proposed UDFs optimized for ML and LLM inference. EvaDB~\cite{aero_2025} offers LLM inference in SQL with UDFs and an adaptive optimization technique to route conjunctive predicates. Flock~\cite{flock_2025} implements an LLM extension for DuckDB with scalar and aggregate LLM inference. However, both lack the logical optimizations which is vital in relational settings. ThalamusDB~\cite{thalamus:2024} support semantic filters and joins where approximate query processing is applied to multi-modal data. 

{\bf Systems with Native Operators:} Few studies propose to integrate model inference at a granular level in the database as native ML operators. Recent work~\cite{rieger_2022} implements a model inference operator in the Umbra database engine that is capable of loading and executing pre-trained models. It aims to minimize data movement and copy. However, it is limited to tabular deep learning models. Similarly, other work~\cite{klabe_2022} implements a deep learning operator that has common primitive layers (fully connected layer, RNN, LSTM) and a model weight storage technique. It has a novel technique for in-database inference. However, its set of compatible models is limited. \cite{konstantinos_2020} extends  towards inferencing deep learning models.

In contrast, \dbname{} combines the strengths of relational databases with flexible LLM inference. It allows scalar, table, and aggregate functions in SQL queries to directly use LLM-based information extraction, filtering, and aggregation on both structured and unstructured data. 
\dbname{}
supports typed data extraction, semantic joins, and compatibility of both local and remote models, resulting in a generalized, declarative, and extensible solution that bridges traditional query execution with semantic operations.

%% file: 10_conclusion.tex
We introduce \dbname{}, an extension of SQL with inline LLM inference, enabling semantic queries directly within relational databases. By unifying relational operations with LLM based semantic operators, \dbname{} enables semantic querying without external pipelines. We demonstrate a series of use case queries on how this integration offers a practical, declarative approach for combining traditional SQL with semantic data processing. The optimizations introduced for logical planning and the execution of semantic operators show up to \improvement{} over the state-of-the-art in terms of latency.
\newpage